\documentclass[letterpaper,aps,prc,superscriptaddress,showpacs,nofootinbib,floatfix,twocolumn]{revtex4}
\usepackage{doi,graphicx,amssymb,amsmath}
\usepackage[usenames,x11names]{xcolor}
\usepackage{txfonts}

\newcommand*{\ee}[1]{\ensuremath{\times 10^{#1}}}
\newcommand*{\sci}[2]{\ensuremath{#1\ee{#2}}}
\newcommand{\power}[2]{\ensuremath{{#1}^{#2}}}
\newcommand*{\unitskip}{\,}
\newcommand*{\unitstyle}[1]{\mathrm{#1}}
\newcommand*{\val}[2]{\ensuremath{#1\unitskip#2}}

\newcommand*{\valrng}[4][~to~]{\ensuremath{\left(#2\textrm{#1}#3\right)#4}}

\newcommand*{\kilo}{\unitstyle{k}}
\newcommand*{\Mega}{\unitstyle{M}}
\newcommand*{\Giga}{\unitstyle{G}}

\newcommand*{\meter}{\unitstyle{m}}

\newcommand*{\second}{\unitstyle{s}}

\newcommand*{\Kelvin}{\unitstyle{K}}
\newcommand*{\K}{\Kelvin}  

\newcommand*{\erg}{\unitstyle{erg}} 

\newcommand*{\ergspersecond}{\erg\unitskip\power{\second}{-1}}

\newcommand*{\amu}{\unitstyle{u}} 
\newcommand*{\fermi}{\unitstyle{fm}} 
\newcommand*{\eV}{\unitstyle{eV}}        
\newcommand*{\MeV}{\Mega\eV} 
\newcommand*{\GeV}{\Giga\eV} 

\newcommand*{\Msun}{\ensuremath{M_\odot}}

\newcommand*{\nuclei}[2]{\ensuremath{\mathrm{^{#1}#2}}}

\newcommand*{\electron}{\ensuremath{\mathrm{e^-}}}
\newcommand*{\ele}{\electron}
\newcommand*{\neutron}{\ensuremath{\mathrm{n}}}
\newcommand*{\nt}{\neutron}
\newcommand*{\proton}{\ensuremath{\mathrm{p}}}
\newcommand*{\pt}{\proton}

\newcommand*{\helium}[1][4]{\nuclei{#1}{He}}

\newcommand*{\beryllium}[1][9]{\nuclei{#1}{Be}}

\newcommand*{\magnesium}[1][24]{\nuclei{#1}{Mg}}

\newcommand*{\calcium}[1][40]{\nuclei{#1}{Ca}}

\newcommand*{\nickel}[1][58]{\nuclei{#1}{Ni}}

\newcommand*{\zirconium}[1][94]{\nuclei{#1}{Zr}}

\newcommand*{\ruthenium}[1][102]{\nuclei{#1}{Ru}}

\newcommand*{\indium}[1][115]{\nuclei{#1}{In}}
\newcommand*{\tin}[1][120]{\nuclei{#1}{Sn}}

\newcommand*{\samarium}[1][152]{\nuclei{#1}{Sm}}

\newcommand*{\lead}[1][208]{\nuclei{#1}{Pb}}

\newcommand*{\Teffinf}{\ensuremath{T_{\mathrm{\!eff},\infty}}}	
\newcommand*{\Tbbinf}{\ensuremath{T_{\mathrm{bb},\infty}}}	
\newcommand*{\MdUrca}{\ensuremath{M_{\mathrm{dUrca}}}}      
\newcommand*{\dens}{\ensuremath{\rho}}  
\newcommand*{\satdens}{\ensuremath{\dens_0}}    
\newcommand*{\eps}{\ensuremath{\epsilon}}  
\newcommand*{\rhoz}{\satdens}

\newcommand*{\alphad}{\ensuremath{\alpha_{\mathrm{D}}}}
\newcommand*{\alphadDM}{\ensuremath{\alpha_{\mathrm{D}}^{\mathrm{DM}}}}

\newcommand*{\RskinPb}{\ensuremath{R_{\mathrm{skin}}^{208}}}

\newcommand*{\rvfigscale}{0.85}
\newlength{\rvcolwidth}
\newlength{\figwidth}
\begin{document}

\title{A way forward in the study of the symmetry energy: experiment, theory, and observation}

\author{C. J. Horowitz}\email{horowit@indiana.edu}
\affiliation{Department of Physics and Nuclear Theory Center, Indiana University, Bloomington, IN 47405, USA}
\author{E. F. Brown}\altaffiliation{Joint Institute for Nuclear Astrophysics}
\affiliation{Department of Physics and Astronomy, and National Superconducting Cyclotron Laboratory, Michigan State University, East Lansing, MI 48824, USA}
\author{Y. Kim}
\affiliation{Rare Isotope Science Project, Institute for Basic Science, Daejeon 305-811, Republic of Korea}
\author{W. G. Lynch}
\affiliation{Department of Physics and Astronomy, and National Superconducting Cyclotron Laboratory, Michigan State 
University, East Lansing, MI 48824, USA}
\author{R. Michaels}
\affiliation{Thomas Jefferson National Accelerator Facility, Newport News, VA, USA}
\author{A. Ono}
\affiliation{Department of Physics, Tohoku University, Sendai 980-8578, Japan}
\author{J. Piekarewicz}
\affiliation{Department of Physics, Florida State University,
               Tallahassee, FL 32306, USA}
\author{M. B. Tsang}
\affiliation{Department of Physics and Astronomy, and National Superconducting Cyclotron Laboratory, Michigan State University, East Lansing, MI 48824, USA}
\author{H. H. Wolter}
\affiliation{Fakult\"at f\"ur Physik, Universit\"at M\"unchen, Am Coulombwall 1, D-85748 Garching, Germany}
\date{\today}

\begin{abstract}
The symmetry energy describes how the energy of nuclear matter rises as one goes away from equal numbers of neutrons and protons.  This is very important to describe neutron rich matter in astrophysics.  This article reviews our knowledge of the symmetry energy from theoretical calculations, nuclear structure measurements, heavy ion collisions, and astronomical observations.  We then present a roadmap to make progress in areas of relevance to the symmetry energy that promotes collaboration between the astrophysics and the nuclear physics communities.
\end{abstract}

\pacs{21.65.Cd, 21.65.Mn, 26.60.-c, 26.50.+x}

\maketitle

\section{Introduction}\label{sec:Intro}

The liquid-drop formula of Bethe and Weizs\"acker\,\cite{Weizsacker:1935,Bethe:1936}
models the nucleus as 
an incompressible quantum drop consisting of $Z$ protons, $N$ neutrons,
and mass number $A = Z + N$. In particular, the nuclear binding energy 
is expressed in terms 
of a handful of empirical parameters that capture the physics of a quantum 
drop. That is,
\begin{equation}
 B(Z,N) = a_{V}A - a_{S}A^{2/3} - 
 a_{C}\frac{Z^{2}}{A^{1/3}} - a_{A}\!\frac{(N-Z)^{2}}{A}+\ldots
 \label{BWMF}
\end{equation}
The volume term $a_{V}$ represents the binding energy per nucleon of 
a large symmetric drop in the absence of long-range Coulomb forces. In turn, 
the next three terms denote binding-energy corrections resulting from the
development of a nuclear surface, the Coulomb repulsion among protons, 
and the Pauli exclusion principle and strong interactions that favor symmetric 
($N = Z$) systems. Although refinements to the mass formula have been 
made to account for the emergence of nuclear shells, the structure of this 75 
year-old formula has remained practically unchanged.

In the thermodynamic limit in which both the number of nucleons and the volume 
are taken to infinity but their ratio remains fixed at the saturation density, the 
binding energy per nucleon may be written as
\begin{equation}
 \eps(\alpha) \equiv -\frac{B(Z,N)}{A} = -a_{V} + J\alpha^{2} \,,
 \label{EAsym}
\end{equation}
where $J \equiv a_{A}$ and $\alpha = (N-Z)/A$ is the neutron-proton 
asymmetry. Note that we have neglected long-range Coulomb forces (which
would render the drop unstable) and have assumed that both $Z$ and $N$ are 
individually conserved. Such a simple expression suggests that the binding
energy per nucleon of a large symmetric drop of density 
$\satdens \approx \val{0.15}{\fermi^{-3}}$ is $a_{V} \approx \val{16}{\MeV}$ and 
that there is an energy cost of $J\approx\val{32}{\MeV}$ in converting all protons 
into neutrons. However, in reality the liquid drop is not incompressible, so the 
semi-empirical mass formula, while highly insightful, fails to describe the 
response of the liquid drop to density fluctuations. This information is contained 
in the equation of state (EOS) which dictates the dependence of the energy per 
nucleon on both the density and the neutron-proton asymmetry. 
Following Eq.~(\ref{EAsym}), we may write the equation of state of asymmetric 
matter as
\begin{equation}
 {\mathcal E}(\rho,\alpha) = {\mathcal E}(\rho,\alpha = 0) + 
 S(\rho)\alpha^{2} + \ldots
 \label{EOS}
\end{equation}
where  ${\mathcal E}(\rho,\alpha = 0)$ is the EOS of symmetric nuclear matter
and $S(\rho)$ is the symmetry energy:
\begin{equation}
 S(\rho) \equiv \frac{1}{2}
 \left(\frac{\partial^{2}{\mathcal E}(\rho,\alpha)}{\partial\alpha^{2}}\right)_{\alpha=0}
 \approx \mathcal{E}(\rho,\alpha = 1)-\mathcal{E}(\rho,\alpha = 0)
 \label{SymmE}.
\end{equation}
Although the value of the symmetry energy at a density of 
$\dens \approx\val{0.1}{\fermi^{-3}}$ is fairly well
constrained by the masses of heavy nuclei, at present its density dependence is 
poorly known. Note that the EOS of asymmetric matter is mainly characterized by 
the density dependence of the symmetry energy ($\partial S/\partial\dens$) which 
is essential for the understanding of the structure of neutron-rich nuclei, particle 
yields in heavy-ion collisions, and properties of neutron stars. To characterize the 
departure of the symmetry energy from its value at saturation, it is customary to 
perform a Taylor series expansion around saturation density. That 
is \cite{Piekarewicz:2008nh},
\begin{equation}
 S(\rho) = J + Lx + \frac{1}{2}K_{\rm sym}x^{2}+\ldots   
\label{JLK}
\end{equation}
where $x = (\rho - \satdens)/3\satdens$. In particular, an enormous effort has been, and 
continues to be, devoted to determine the slope of the symmetry energy $L$,
\begin{equation}
L\equiv 3\satdens\left.\left(\frac{\partial S}{\partial \dens}\right)\right|_{\satdens} .
\label{Eq.defL}
\end{equation} 
The nuclear symmetry energy describes the increase in energy as matter changes away 
from a symmetric configuration of equal numbers of neutrons and protons (or up and down 
quarks). Observables involving moderately neutron-rich nuclei up to very neutron-rich 
astrophysical systems are highly sensitive to the density dependence of the symmetry 
energy. Interpretation of these observables is often hindered, however, by uncontrolled 
extrapolations. Thus the need for a systematic and comprehensive program to determine 
the density dependence of the symmetry energy  becomes critical \citep[see Refs.][and 
references contained therein]{Steiner:2004fi,tsang12}. The aim of this contribution is to 
foster dialogue among the nuclear and astrophysical communities on how best 
to achieve this common goal.

Ideally, one would like to calculate the density dependence of the symmetry energy starting 
from QCD. Indeed, dense QCD is intimately connected to nuclear physics and astrophysics
and offers a fertile testing ground for our understanding of quantum-field theories in the 
non-perturbative regime. Moreover, the rich phase diagram of baryonic matter is believed to
exhibit unique and novel states of matter that should be directly predicted by QCD. 
Unfortunately, at present no theoretical framework is available to study dense matter from 
first principles. Whereas lattice QCD has been very successful in simulating the thermal 
component of the EOS, progress in describing matter at finite baryon density has been 
exceedingly slow. Thus, in practice one must rely on phenomenological approaches 
calibrated from both laboratory experiments and astrophysical observations. As such,
existing and forthcoming rare-isotope facilities will play an essential role in elucidating the 
nature of the phase diagram of strongly interacting matter and will provide critical inputs for refining the theoretical models of dense matter. 

The density dependence of the symmetry energy plays a critical role in shaping the structure 
of finite nuclei. In particular, the neutron-skin thickness of heavy nuclei is highly sensitive to
the difference between the symmetry energy at saturation density (as in the nuclear core) and 
the symmetry energy at lower densities (as in the nuclear surface). Ultimately, the thickness
of the neutron skin emerges from a dynamical competition between the surface tension and 
the slope of the symmetry energy $L$. Moreover, the electric dipole
polarizability, described below, is also highly
sensitive to the density dependence of the symmetry energy. The dipole polarizability is an
ideal complement to the neutron-skin thickness because the symmetry energy acts as the 
restoring force. 

The density dependence of the symmetry energy also has a profound impact on a variety of 
astrophysical phenomena. At very low densities, uniform neutron-rich matter becomes 
unstable against cluster formation. The onset of the instability and the formation of clusters 
is controlled by the symmetry energy; this may be important during the collapse of the core 
of a massive star or during the merger of two neutron stars. As such, the symmetry energy 
may play an important role in setting the conditions under which r-process nucleosynthesis 
occurs. Moreover, for densities of $\dens\approx\valrng[--]{2}{3}{\satdens}$ a significant 
component of the pressure of neutron-rich matter is determined by $\partial S/\partial\dens$, 
and this in turn determines the neutron-star radius. Finally, at even higher densities, the 
symmetry energy controls the proton fraction, which in turn dictates whether enhanced
neutrino cooling of neutron stars, via the direct Urca process is 
possible \citep{yakovlev.pethick:neutron}, see Sec.~\ref{subsec:NScooling}. 

In the laboratory, the density dependence of the symmetry energy can be probed via heavy-ion 
collisions at different beam energies and with nuclei having a wide range of neutron-proton 
asymmetries. Indeed, heavy-ion collisions provide the only means to study the EOS of 
asymmetric matter under controlled laboratory conditions. In particular, low-energy heavy-ion 
collisions can produce warm, dilute neutron-rich matter that closely resemble the conditions in 
the neutrinosphere (i.e., the surface of last neutrino scattering) in a core-collapse supernova.
At higher beam energies, collisions of neutron-rich heavy ions may shed light on the symmetry 
energy at densities of $\dens \approx 2{\satdens}$, a region that is critical in the development 
of the neutron-star radius. At these energies, pion production and particle flow may serve as 
useful probes of the symmetry energy at high densities. Critical to the success of this endeavor, 
however, is the reliability of transport models in distilling details of the symmetry energy from 
heavy-ion observables.

The purpose of this article is to provide a roadmap for future progress in studying the symmetry energy. 
After the brief motivation presented in this Introduction, we continue by discussing recent progress and 
ongoing efforts in a variety of topics of relevance to the symmetry energy, such as microscopic calculations, 
dilute neutron-rich matter, nuclear structure, neutron-star matter, and heavy-ion collisions. We then conclude 
with a description of our vision for the ``way forward'' in each of these areas.

\section{Calculations of the symmetry energy}

Our inability to calculate the symmetry energy from first principles increases the importance of both laboratory
experiments and astronomical observations. One can, however, calculate $S$ from first principles in the high 
temperature, low density regime. Furthermore, chiral effective field theory provides promising results for $S$ at 
$\dens\lesssim\satdens$. At higher densities, a variety of calculations exist but these rely on mostly phenomenological 
approaches. In what follows, we review calculations of the symmetry energy at high temperatures in 
Sec.~\ref{subsec:highT}, at densities near \satdens\ in Sec.~\ref{subsec:n0}, and finally at high densities in 
Sec.~\ref{subsec:highn}.

\subsection{Symmetry energy at high temperatures and low densities}
\label{subsec:highT}
At temperatures $T\gtrsim \MeV$ and densities $\dens\ll\satdens$, one can calculate $S$ 
exactly using a virial expansion\,\cite{Horowitz2006}.  In this limit the equation of state may be expanded in 
powers of the fugacity $z_i=\exp(\mu_i/T)$, where $\mu_i$ is the chemical potential for species $i$. The virial 
expansion is valid near the classical limit of $z_i \ll 1$. Quadratic coefficients of the virial expansion are 
calculated from nucleon--nucleon, nucleon--light-cluster, and light-cluster--light-cluster elastic scattering phase 
shifts. Note that there are important contributions from alpha particles that can significantly increase the 
symmetry energy. The symmetry energy computed from the virial expansion provides an exact critical 
benchmark for supernova simulations (see Sec.~\ref{sec:verylowdensity}) that can be probed in the laboratory 
with heavy-ion collisions; \citep[see, for example, Ref.][]{Natowitz2010}.  At temperatures $\gtrsim\val{50}{\MeV}$ or 
above, one can extend the virial expansion to also include pion degrees of freedom. At low densities and even 
higher temperatures $T\gtrsim \val{100}{\MeV}$, where $\mu_i/T \ll 1$, one can calculate $S$ directly from QCD simulations. In general, lattice QCD can calculate the 
equation of state for any temperature---in the limit of zero baryon density. Although lattice simulations at arbitrary 
densities are hindered by the conspicuous ``sign problem,'' progress in this area has been achieved by expanding 
around the zero baryon density limit. Indeed, if the EOS is expanded in powers of the chemical potential to the
temperature, lattice QCD simulations have been successful in calculating the relevant coefficients through order
$(\mu_i/T)^4$ \cite{Borsanyi:2011sw,Bazavov:2012jq}. In a future work we will use these coefficients to 
constrain the symmetry energy at high temperatures. Although symmetry energy effects at high temperatures 
may be small relative to the large thermal energies, these model-independent QCD results provide important 
benchmarks in the calibration of theoretical models.

\subsection{Symmetry energy near saturation density}
\label{subsec:n0}
Chiral effective field theory is a powerful approach in which nuclear interactions are systematically expanded in 
powers of the momentum transfer over a typical chiral (e.g., pion mass) scale. Perhaps the most important 
feature of the chiral approach is its hierarchical nature. That is, provided that the chiral expansion converges,  
two-nucleon forces dominate over three-nucleon forces, which in turn are more important than four-nucleon forces, 
and so on. This allows one to perform many-body calculations for which the complexity of the Hamiltonian is 
manageable. Recently, \citet{Kruger2013} used many-body perturbation theory to calculate the energy of infinite 
neutron matter to order N3LO by including two-, three-, and four-nucleon forces. Moreover, given the systematic
nature of the expansion, estimates of the theoretical uncertainties were also provided. To provide estimates for
both the symmetry energy $J$ and slope $L$ at saturation density, they approximate the symmetry energy 
[as in Eq.~(\ref{SymmE})] as the difference between the energy of pure neutron matter---which they 
calculate---and the energy of symmetric nuclear matter, which they do not calculate but instead adopt the 
empirical saturation point.   The following values were reported \citep{Kruger2013}: 
$S_0 \equiv J = \valrng[--]{28.9}{34.9}{\MeV}$ and  $L = \valrng[--]{43}{67}{\MeV}$. However, note that
\citeauthor{Kruger2013} do not calculate the EOS of pure neutron matter much beyond saturations density,
as the chiral expansion may converge poorly at higher densities.

\subsection{Symmetry energy at high densities}
\label{subsec:highn}
At present neither laboratory experiments nor astronomical observations place stringent constraints on 
the symmetry energy at high densities.  Calculations of $S$ at high densities are hindered by large 
uncertainties related to the poor convergence of the chiral expansion. For example, \citet{Gandolfi2013} 
calculate the energy of pure neutron matter up to densities of $\dens\gtrsim 3\satdens$ using Quantum 
Monte Carlo techniques with phenomenological two- and three-nucleon forces. In particular, by varying the 
three-nucleon force, they find a sharp linear correlation between $S_0$ and $L$. However, without proper
theoretical guidance, for example from the chiral approach, it is unclear whether the form of their 
three-nucleon force is reliable. Moreover, without a proper expansion (chiral or otherwise) it is also unclear 
whether four-nucleon (or higher order) forces could make a significant contribution at high densities. Thus, 
it is likely that these calculations may have large theoretical uncertainties at high densities. Note that there 
are also calculations of the symmetry energy using Brueckner many-body theory with phenomenological 
two- and three-nucleon forces, see for example \cite{PhysRevC.88.035805,PhysRevC.84.044307}. However, these calculations may also suffer from large uncertainties related to
their phenomenological Hamiltonians and the role of many-nucleon forces.

In summary, chiral effective field theory provides a promising way to calculate the symmetry energy for uniform matter at low densities.   At very low densities, matter is likely nonuniform and this must be taken into account when calculating the symmetry energy.  One way to do this is by including clusters in a virial expansion.  Unfortunately, all present calculations of the symmetry energy at high densities (above nuclear saturation) may have large uncertainties related to the form of the interactions.  Therefore one may need to rely instead on phenomenological models, astrophysical observations, and heavy ion experiments.  For example Refs.~\citenum{Lee:2010sw,Dong:2012ch} and \citenum{Kim:2010dp} describe two very different phenomenological models of $S$ at high densities.

\section{Symmetry energy at very low densities in nonuniform matter}
\label{sec:verylowdensity}
At low densities, uniform nuclear matter becomes unstable against cluster formation. Indeed, 
at densities of $\dens\lesssim\satdens/2$ the inter-nucleon separation becomes comparable 
to the range of the nucleon-nucleon interaction, so it becomes energetically favorable for the 
system to fragment into neutron-rich clusters. Cluster formation significantly increases the 
symmetry energy at very low densities and this may be of relevance to the modeling of core 
collapse supernovae (CCSN). These giant stellar explosions radiate away the large gravitational 
binding energy of a neutron star ($\sim\val{100}{\MeV/\textrm{nucleon}}$) by emitting $\approx 10^{58}$ neutrinos. 
Much of the ``action" in CCSN happens near the neutrinosphere, which defines the transition region
between interacting and free-streaming neutrinos. The neutrinosphere is composed of a warm 
low-density gas of neutron-rich matter at temperatures of $T \sim \val{5}{\MeV}$ and densities 
of $\dens\sim\val{0.01}{\satdens}$. Here the neutrino mean free path becomes comparable to the size 
of the system. 

Recently, several groups have come to appreciate that in addition to a free gas of neutrons and 
protons, the neutrinosphere may also contain light nuclei \citep{Horowitz2006,Arcones2008} and 
may display important many-body correlations \citep{Roberts2012}. These many-body effects 
can have a significant impact on neutrino transport by modifying the neutrino opacity and 
ultimately the emitted neutrino spectra \citep{Roberts2012a,Pinedo2012}. In particular, this may 
be important for nucleosynthesis in the innermost regions of core collapse supernovae.

Remarkably, many properties of the neutrinosphere can be directly reproduced in the laboratory 
with heavy-ion collisions. Temperatures of about $\val{5}{\MeV}$ are easy to achieve, while low subnuclear 
densities can be studied, for example, by observing intermediate velocity fragments from peripheral 
collisions \cite{Natowitz2010}.  Perhaps the most difficult property to simulate in these collisions is 
the large neutron-to-proton ratio present in the neutrinosphere. However, new and forthcoming radioactive 
beam facilities, such as RIKEN (already in operation) and FRIB (planned for 2020), allow heavy-ion
collisions involving systems with widely varying neutron-proton asymmetries. By comparing results 
between neutron-deficient and neutron-rich systems, one should be able to reliably extrapolate to systems 
with extreme neutron excess; see Sec.~\ref{sec:HI}. By measuring the composition of light clusters,
these laboratory experiments should be able to infer the symmetry energy and the corresponding 
EOS of asymmetric nuclear matter at very low density. In turn, these measurements should 
provide critical inputs for new microscopic approaches to warm, dilute matter that face the challenge 
of accounting for many-body correlations and for the concomitant development of nuclear clusters.




\section{Symmetry Energy and Nuclear Structure}
\label{sec:nuclear_structure}

\subsection{Neutron Skins}
\label{subsec:neutronskins}

Starting with the pioneering work of Hofstadter in the late 1950's \cite{Hofstadter:1956qs} 
and continuing to this day, measurements of charge distributions of nuclei using elastic 
electron scattering have provided knowledge of charge radii with remarkable accuracy 
across the nuclear chart\,\cite{Fricke:1995,Angeli:2013}. In contrast, probing neutron 
densities has traditionally relied on hadronic experiments that are hindered by large and 
uncontrolled uncertainties. Recently, the Lead Radius Experiment (PREX) at the Thomas 
Jefferson National Accelerator Facility (Jefferson Lab) has pioneered parity-violating 
measurements of neutron radii by relying on the significantly larger weak charge of the 
neutron relative to that of the proton\,\cite{Abrahamyan:2012gp,Horowitz:2012tj}. 
Measurements of the parity-violating asymmetry $A_{PV}$ at Jefferson Lab using 
longitudinally polarized electrons is an established technique that has been used 
successfully to probe the quark structure of the nucleon\cite{ref:happex-1,ref:happex-2,
ref:happex-3,ref:happex-4,ref:happex-5,G0sff-1,G0sff-2,G0sff-3,Androic:2013rhu}.
The parity-violating asymmetry is defined as the difference in the cross section between
right- and left-handed longitudinally polarized electrons relative to their sum. That is,
\begin{equation}
  A_{PV} = \frac{\sigma_R - \sigma_L}{\sigma_R + \sigma_L}\,.
  \label{APV}
\end{equation}
This powerful technique provides a unique opportunity to measure the weak charge 
form factor of the nucleus---and hence its neutron radius $R_n$---in a relatively clean 
and model-independent way\,\cite{dds,bigprex,RocaMaza:2011pm,Reinhard:2013fpa}. 
Indeed, PREX has provided for the first 
time model-independent evidence---at the $1.8\sigma$ level---in favor of a neutron-rich 
skin in \lead[208] and successfully demonstrated the feasibility of this technique for 
measuring neutron densities with an excellent control of systematic 
errors\,\cite{Abrahamyan:2012gp,Horowitz:2012tj}. The neutron-skin thickness, defined 
as the difference between the neutron ($R_n$) and proton ($R_p$) root-mean-square radii 
was reported to be
\begin{equation}
 \RskinPb = R_{n}^{208}-R_{p}^{208}=\!{0.33}^{+0.16}_{-0.18}\,{\rm fm}.
\end{equation}
In a follow-up already approved experiment ``PREX-II'',  the uncertainty in the 
determination of $R_{n}^{208}$ will be reduced by a factor of three,
to $\pm\val{0.06}{\fermi}$. 

The neutron-skin thickness of \lead[208] is strongly sensitive to the density dependence 
of the symmetry energy. In particular, $\RskinPb$ is strongly correlated to the 
slope of the symmetry energy $L$\,\cite{Centelles:2008vu}. 
This correlation is strong because it emerges from 
simple, yet robust, physical arguments. In the spirit of the liquid-drop model, surface 
tension favors the formation of a spherical drop of uniform equilibrium density. However, 
for a neutron-rich system it is unclear whether the extra neutrons should reside in the 
surface or in the core. Placing them in the core is favored by surface tension but 
disfavored by the symmetry energy, which is large at saturation density. Conversely, 
moving them to the surface increases the surface tension but reduces the symmetry 
energy. Hence, the neutron-rich skin of a heavy nucleus emerges from a dynamic 
competition between the surface tension and the {\sl difference} between the symmetry 
energy at saturation density and at a lower surface density. In particular, for a stiff 
symmetry energy, namely one that increases rapidly with density, it is energetically 
favorable to move most of the the neutrons to the surface where the symmetry energy 
is low; this generates a thick neutron skin. 

\begin{figure}[htbp]
\centering
  \includegraphics[width=\figwidth]{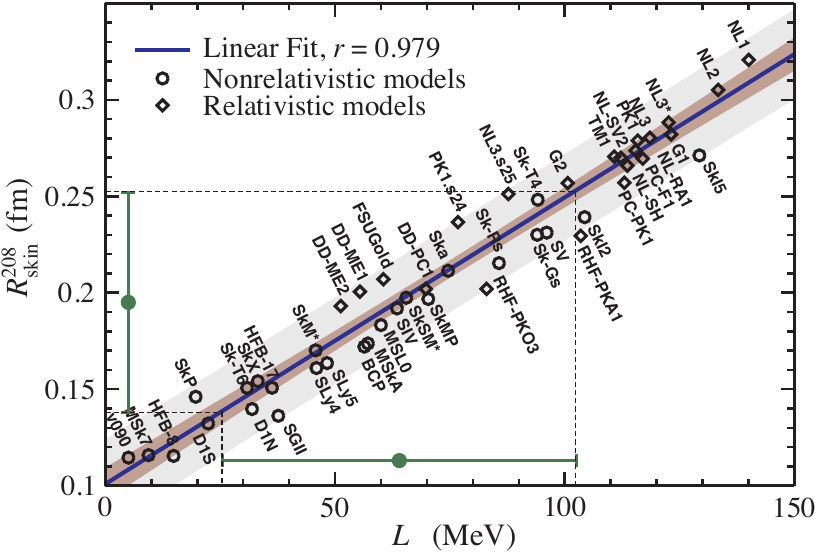}
  \caption{(Color online) Predictions from a large number of EDFs for the neutron-skin thickness of \lead[208], \RskinPb, as a function of the slope of the symmetry energy $L$. The thicker shaded region represents the 95\% prediction band of the linear regression. The error bar in $L$ is derived by assuming a $\pm\val{0.06}{\fermi}$ error in the measurement of $\RskinPb$ (around an arbitrary central value).  These results were first reported in Ref.~\citenum{RocaMaza:2011pm}, and the figure is adapted from that reference.
  }
 \label{FigNSkins}
\end{figure}

Given that at zero temperature the  difference between the symmetry energy at two 
neighboring density points is directly proportional to the symmetry pressure $L$, a 
strong correlation between $\RskinPb$ and $L$ is expected. Indeed, in 
Fig.~\ref{FigNSkins} we display the strong correlation (with a correlation coefficient
of $r=0.979$) between $\RskinPb$ and $L$ using a large and representative 
set of Energy Density Functionals (EDFs); note that this figure was first published 
by\,\citet{RocaMaza:2011pm}. A linear fit to the predictions of all the models displayed 
in Fig.~\ref{FigNSkins} yields:
\begin{equation}
 \RskinPb=\frac{r_{\rm s}}{2}\left(\frac{L+L_{\rm s}\pm\delta L_{\rm s}}{L_{\rm s}}\right) ,
 \label{R208vsL}
\end{equation}
where the three fitting parameters are given by $r_{\rm s}=\val{0.2}{\fermi}$, $L_{\rm s}=\val{68.7}{\MeV}$, 
and $\delta L_{\rm s}=\val{6.8}{\MeV}$ represents the 70\% prediction-band error. In particular, this suggests 
that a $\pm\val{0.06}{\fermi}$ error in the measurement of $\RskinPb$ (as indicated in Fig.\,\ref{FigNSkins})
would translate into a $1\sigma$ error in $L$ of $\Delta L(\val{0.06}{\fermi})=\val{40.8}{\MeV}$. To properly estimate 
the final uncertainty in the determination of $L$, one should add both theoretical and experimental errors in quadrature.

\subsection{Isoscalar Monopole Resonance}
The isoscalar monopole resonance measures the collective response of the nucleus 
to density variations. Pictorially, this collective excitation in which protons and 
neutrons oscillate in phase around the equilibrium density may be perceived as a 
nuclear {\sl breathing} mode. Given that symmetric nuclear matter saturates, the 
pressure at saturation density vanishes. Thus, the giant monopole resonance (GMR)
probes the curvature of the Equation of State (EOS) at saturation density, or equivalently, the 
incompressibility coefficient of symmetric nuclear matter $K_{0}$. However, the 
accurate determination of $K_{0}$ requires the formation of a strong collective peak 
that involves many nucleons and exhausts most of the energy weighted sum rule. 
Thus, the coherent response of the system is sensitive to the  incompressibility 
coefficient of neutron-rich matter $K_{0}(\alpha)$ rather than to only $K_{0}$;
here $\alpha\!=\!(N\!-\!Z)/A$ is the neutron-proton asymmetry. As such, $K_{0}(\alpha)$ 
is sensitive to the density dependence of the symmetry energy. Indeed, the 
incompressibility coefficient of asymmetric nuclear matter may be written as 
follows\cite{Piekarewicz:2008nh}:
\begin{equation}
  K_{0}(\alpha)=K_{0}+K_{\tau}\alpha^{2}
            \equiv K_{0}+\Big(K_{\rm sym}-6L-\!\frac{Q_{0}}{K_{0}}L\Big)\,\alpha^{2} \;,
\label{Kalpha}            
\end{equation}
where $Q_{0}$ is the skewness parameter of symmetric nuclear matter and
$K_{\rm sym}$ is the curvature of the symmetry energy. 
In principle, to constrain both $K_{0}$ and $K_{\tau}$ one would measure the 
distribution of monopole strength for nuclei with significantly different values 
of $\alpha$. In practice, however, access to the symmetry energy is hindered 
by the relatively low neutron-proton asymmetry of stable nuclei. Nevertheless,
in an effort to determine the incompressibility of neutron-rich matter, a pioneering 
experiment was carried out at the Research Center for Nuclear Physics (RCNP) 
in Osaka, Japan\,\cite{Li:2007bp,Li:2010kfa}. The experiment succeeded in 
measuring the distribution of isoscalar monopole strength in all stable Tin
isotopes having an even number of neutrons, namely, from \tin[112] to \tin[124].
Although the neutron-proton asymmetry along this isotopic chain varies from 
$\alpha=0.11\textrm{--}0.19$, the sensitivity to $K_{\tau}$ is poor---even 
for neutron-rich \tin[124]. Despite this drawback, the experiment uncovered a 
puzzle that remains unsolved after more than 5 years: 
\emph{``Why is Tin so soft?"}\,\cite{Piekarewicz:2007us}. That is, why do models 
that successfully reproduce GMR energies in \zirconium[90], \samarium[144], 
and \lead[208], overestimate the corresponding centroid energies along the 
full isotopic chain in \tin[]. Note that the softness of Tin has been recently 
confirmed in the nearby isotopic chain in Cadmium\,\cite{Patel:2012zd}.

\subsection{Electric Dipole Polarizability}
The oldest known and perhaps most prominent collective nuclear excitation is the isovector
giant dipole resonance (GDR). This mode of excitation is perceived as an 
out-of-phase oscillation of neutrons against protons. Given that this oscillation 
results in the separation of two dilute quantum fluids---one neutron-rich and the 
other one proton-rich---the symmetry energy acts as the restoring force. 
The GDR is one state that contributes to the electric dipole polarizability.
By using a covariance analysis with an accurately-calibrated density functional,
it was recently demonstrated that the electric dipole polarizability $\alphad$ 
is a strong isovector indicator that is highly correlated to the neutron-skin
thickness of heavy nuclei\,\cite{Reinhard:2010wz}. Shortly after, using a large 
number of EDFs, it was confirmed that such a correlation is robust---although 
some systematic model dependence emerged\,\cite{Piekarewicz:2012pp}. 

\begin{figure*}[htbp]
 \begin{center}
  \includegraphics[width=\figwidth]{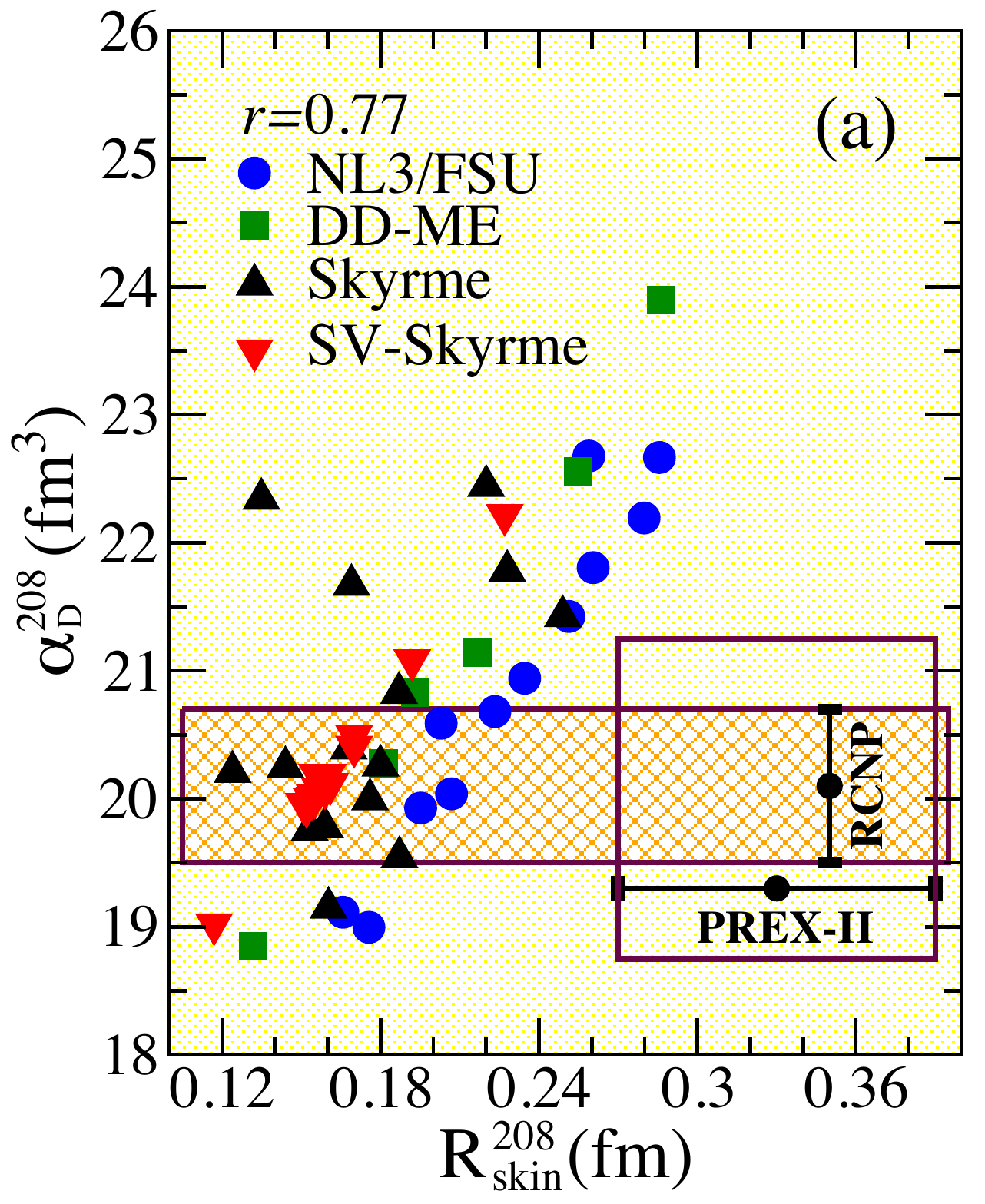}
 	\hspace{0.05\textwidth}
  \includegraphics[width=\figwidth]{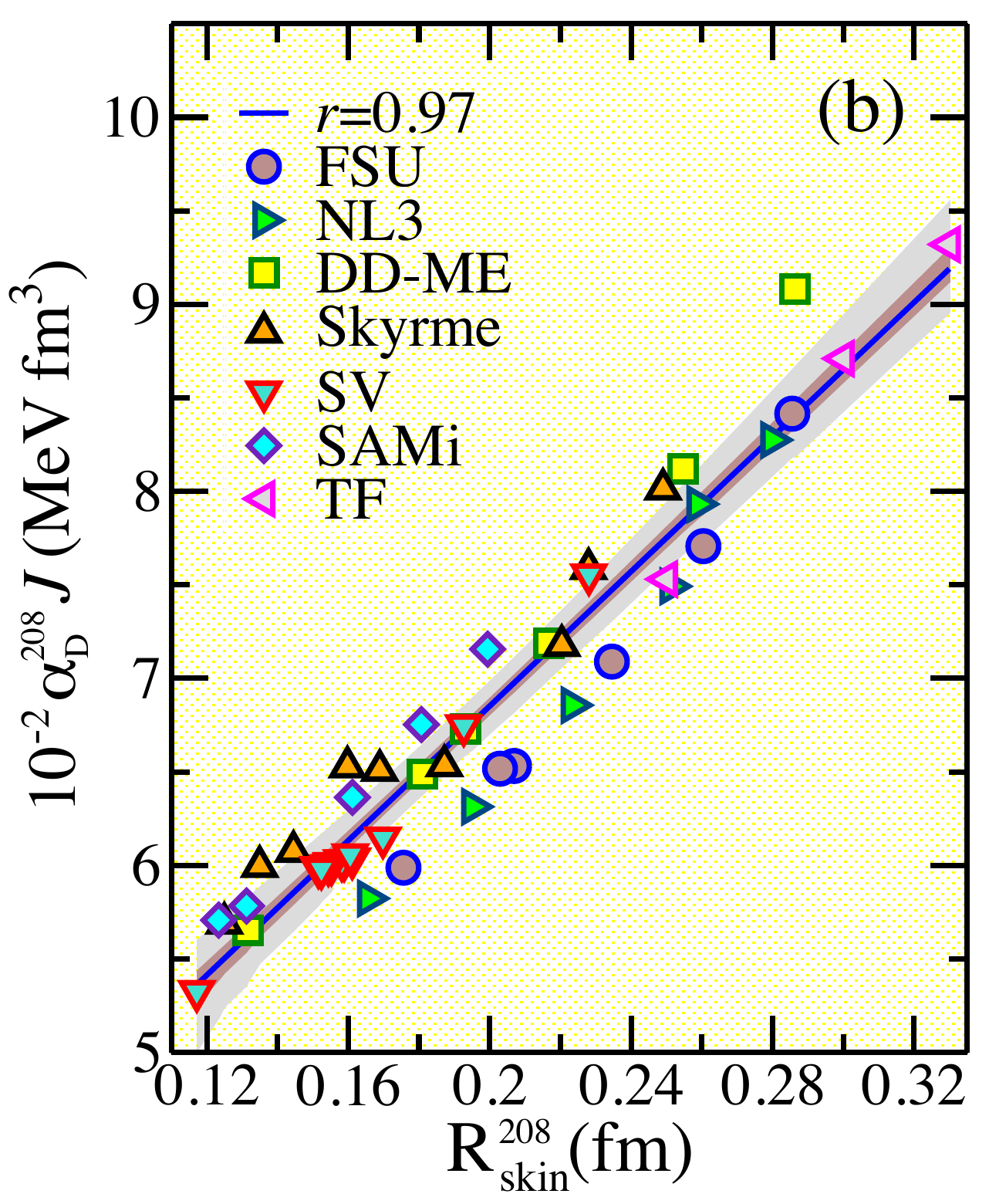}
  \caption{(Color online) 
  (a) Predictions from a large number of EDFs for the electric dipole polarizability and 
  neutron-skin thickness of \lead[208] as discussed in Ref.\,\cite{Piekarewicz:2012pp}. 
  Constrains from both RCNP\,\cite{Tamii:2011pv,Poltoratska:2012nf} and PREX\,\cite{Abrahamyan:2012gp} 
  (the latter assuming a projected 0.06\,fm error) have been incorporated into the plot. 
  (b) Predictions from a large number of EDFs for $\alphad^{208}$ times the symmetry energy
  at saturation density $J$ as a function of $\RskinPb$. The lightly and darkly shaded regions in panel (b)
  represent the 99.9\% and 70\% confidence bands, respectively. These results were first 
  reported in Ref.\,\cite{Roca-Maza:2013mla}.}
 \label{FigIvGDR}
 \end{center} 
\end{figure*}

In an effort to elucidate the connection between the dipole polarizability and the 
density dependence of the symmetry energy, insights from the macroscopic liquid 
droplet model are particularly helpful\,\cite{Satula:2005hy,Roca-Maza:2013mla}. 
Within a droplet model (``DM'') approach, the electric dipole polarizability takes
the following simple form:
\begin{equation}
\alphadDM \approx \frac{\pi e^{2}}{54} \frac{A \langle r^2\rangle}{J}
\left[1+\frac{5}{3}\frac{L}{J}\epsilon_{A}\right]\;,
\label{dpdm}
\end{equation}
where $\langle r^2\rangle$ is the mean-square radius of the nucleus and  
$\epsilon_{A}=(\rhoz-\dens_{A})/3\rhoz$ accounts for the difference between the 
saturation density $\rhoz$ and an appropriate average density of the nucleus $A$ is $\dens_{A}$. 
To show the value of this qualitative formula in the particular case of \lead[208], we 
display on the left-hand panel of Fig.\,\ref{FigIvGDR} the correlation between $\alphad^{208}$ 
and $\RskinPb$ as predicted by a large number of EDFs~\cite{Piekarewicz:2012pp}. 
Although a clear correlation between the dipole polarizability and the neutron-skin thickness 
is observed, there is an appreciable scatter in the predictions; this yields a correlation 
coefficient of $r=0.77$. Note that by imposing the recent experimental constraint from 
$\alphad^{208}$ \cite{Tamii:2011pv,Poltoratska:2012nf} several models, especially 
those with a very stiff symmetry energy, may already be ruled out. Remarkably, if the 
follow-up PREX experiment (PREX-II)---with a projected uncertainty of \val{0.06}{\fermi}---finds 
that its central value of $\RskinPb=\val{0.33}{\fermi}$ remains intact, then all models displayed 
in the figure will be ruled out! \cite{Fattoyev:2013yaa}. 
Although the correlation is clear, Eq.~(\ref{dpdm}) suggests that 
the dipole polarizability \emph{times} the symmetry energy at saturation density ($\alphad J$) 
should be far better correlated to $L$ (or equivalently to $\RskinPb$) than $\alphad$ 
alone \cite{Roca-Maza:2013mla}. Thus, the right-hand panel of Fig.~\ref{FigIvGDR} displays 
$J\alphad^{208}$ as a function of $\RskinPb$ as predicted by an augmented set of 
EDFs \cite{Roca-Maza:2013mla}. Remarkably, the fairly large spread in the model predictions 
has been practically eliminated by simply scaling $\alphad^{208}$ by the $J$ associated to 
each model. With a correlation coefficient of $r=0.97$, this suggests that accurate measurements 
of both $\RskinPb$ and $\alphad^{208}$ may provide stringent constraints on $J$ and $L$.

Finally, we note that novel experimental techniques now provide access to other nuclear 
excitations---such as isovector-quadrupole and spin-dipole resonances---that appear to be
sensitive to the density dependence of the symmetry energy. Due to space limitations these 
modes are not discussed here any further; we refer the reader to Ref.\,\cite{Colo:2013yta} 
and references contained therein.

\section{Symmetry energy at high densities in neutron stars}
\label{sec:high_density}
The structure of spherical neutron stars in hydrostatic equilibrium is uniquely 
determined by the EOS of neutron rich matter,
namely, the pressure as a function of energy density. The EOS depends on the 
interactions between nucleons and impacts 
nuclear-structure and heavy-ion observables. Therefore, a measurement in one
domain---be it astrophysics or in the 
laboratory---has important implications in the other.

\subsection{Neutron star radii}
\label{subsec:NSradii}
Observations of neutron star masses and radii directly constrain the EOS. Indeed, a 
complete determination of the 
\emph{Mass-vs-Radius} relationship will uniquely constrain the EOS~\citep{Lindblom:1992}. Critically important to this
task is the recent observation of a two solar mass neutron star by 
\citet{Demorest2010A-two-solar-mas}. The 
determination of the stellar mass is very clean and accurate as it depends on the 
measurement of Shapiro delay in 
the pulsar's radio signal by the gravitational well of its white dwarf companion. 
Furthermore, the existence of 
massive neutron stars has now been confirmed with the discovery of a second two solar 
mass neutron 
star \citep{Antoniadis2013A-Massive-Pulsa}. These observations have immediate and 
important implications:
the pressure of neutron rich matter at high densities (a few times \satdens) must be 
large in order to support a 
two solar mass neutron star against the gravitational collapse into a black hole. In 
particular these observations 
rule out many soft (low pressure) EOS---including those containing hyperons or nearly 
free quarks and gluons. 
However, neutron stars could still harbor quark matter or hyperons in their cores 
provided the (poorly constrained) 
strong interactions among these exotic constituents can significantly increase the 
pressure at high densities.

Combined mass and radius estimates for several neutron stars have been inferred from 
observations of X-ray 
bursts \citep{Ozel2009The-Mass-and-Ra,Guver2010The-Mass-and-Ra,Steiner2010The-Equation-of,Suleimanov2011A-Neutron-Star-,Guver2012Systematic-Unce-1,Guver2012Systematic-Unce-2}. X-ray bursts are thermonuclear explosion triggered 
by the 
accretion of matter from a companion unto the surface of the neutron star. In some 
bursts the luminosity becomes large enough to reach the Eddington limit, in which the  
photosphere is pushed outward from the radiative momentum flux. Thus, at the moment after the burst peak when the photosphere ``touches down,'' the flux observed by a remote observer at a distance $D$ should be
\begin{equation}\label{e.touchdown-flux}
	F_{\mathrm{Edd},\infty} = \frac{GMc}{\kappa D^{2}}\left[1-2\beta(R)\right]^{1/2}.
\end{equation}
Here $\beta(R) = GM/Rc^{2}$ is the compactness, $M$ is the gravitational mass, $R$ is the local radius, i.e., $R = (2\pi)^{-1}\times\{\textrm{circumference of equipotential surface}\}$, and $\kappa$ is the opacity, which depends on the composition of the photosphere. The subscript ``$\infty$'' denotes quantities measured by a distant observer.

In the latter, cooling, stage of the burst, the ratio of the observed flux $F_{\infty}$ to a fitted peak temperature $\Tbbinf$ is roughly constant. This ratio is therefore identified with the observed angular area of the neutron star,
\begin{equation}\label{e.quiescent-angular-area}
\frac{F_{\infty}}{\sigma \Teffinf^{4}} = \left(\frac{R}{D}\right)^{2} 
	\left[1-2\beta(R)\right]^{-1}.
\end{equation}
Here $\sigma$ is the Stefan-Boltzmann constant, and the observed effective temperature $\Teffinf$ is related to $\Tbbinf$ via a composition-dependent correction factor $f_{c}$.

Observations of both 
photospheric expansion and thermal cooling emission from a given system therefore allow a 
separate determination of both $R$ and $M$ from Eqn.~(\ref{e.touchdown-flux})-(\ref{e.quiescent-angular-area}).  
Although in principle this is a powerful technique, there 
are important complications that may
hinder the reliability of the method. For example, the correction $f_{c}$ required to model stellar atmospheres are large, presently uncertain, 
and may vary between different sources. Moreover, it is not yet clear from which X-ray 
burst systems, and at what time during the 
burst, one can extract the most accurate radii, nor is it entirely certain that the radii sampled by both observations are identical \citep{Steiner2010The-Equation-of}. Improved models of X-ray bursts\,
\cite[e.g.,][]{Zamfir2012Constraints-on-} may be 
able to address these uncertainties and allow for a more reliable determination of 
stellar radii.

Perhaps at present the cleanest way to extract neutron-star radii is from the 
observation of quiescent low mass X-ray binaries 
(qLMXB) in globular clusters. Such neutron stars are assumed to have accreted matter 
from a companion in the past, but are 
currently in a quiescent phase where they simply ``glow'' in soft X-rays as they cool.
Often they are assumed to have hydrogen atmospheres for which there are 
very accurate calculations of the emergent spectrum.  The main argument in support of 
this assertion is that other heavier 
elements will rapidly sink in the very strong gravitational field and ultimately leave 
behind a hydrogen atmosphere\,\cite{bildsten92}.  
Moreover, there is no evidence that these neutron stars are strongly magnetized, which 
makes the spectral models simpler.  As a result, the observed effective temperature $T_{\mathrm{eff},\infty}$ can be determined directly by spectral fitting, which removes the systematic uncertainty in $f_{c}$, and the observed angular area can then be determined from Eq.~(\ref{e.quiescent-angular-area}).
Although by itself this technique does not 
allow for independent determinations of $M$ and $R$ for a single source, 
observations of multiple stars---when combined with 
a constraint such as a functional form for the EOS\,\cite{Steiner2013The-Neutron-Sta} or an assumed dependence for 
$R(M)$\,\cite{Guillot2013Measurement-of-}, permits the extraction of both. Although these measurements contain the distance $D$, one can use X-ray binaries in globular clusters---dense collections of up to a million stars---that 
have relatively well-established distances.

Recently \citet{Guillot2013Measurement-of-} determined neutron-star radii from 
fitting the spectra of five qLMXBs with 
models appropriate for a nonmagnetic hydrogen atmosphere. Furthermore, they assumed 
that neutron-radii are approximately 
independent of mass, namely, $R(M) = R_{0}$. Indeed, such an assumption is 
consistent with many---although not 
all---models of the EOS over a wide range of masses \cite{Lattimer:2006xb}. By adopting such an assumption, 
\citet{Guillot2013Measurement-of-} were
able to combine data sets for five different sources and solve for a single neutron-star radius, thereby greatly increasing the 
statistical power of the data. \citet{Guillot2013Measurement-of-} determined that
\begin{equation}
 R_{0}=9.1^{+1.3}_{-1.5}\,{\rm km},
\label{R_bob}
\end{equation}
where the quoted errors are for the 90\% confidence level. This fitting does not account for causality nor pulsar observations indicating that the maximum neutron star mass is greater than approximately $\val{2.0}{\Msun}$ \citep{Demorest2010A-two-solar-mas,Antoniadis2013A-Massive-Pulsa}; imposing these conditions as priors may change the result \citep{Lattimer2013Neutron-Star-Ma}. With these caveats, however, 
Eq.~(\ref{R_bob}) is intriguing 
as it suggests stellar radii significantly 
smaller than those predicted by many theoretical models; at least models with non-exotic constituents \cite{Lattimer:2006xb}.  
Moreover, such a small radius coupled with the observation of neutron 
stars with $M \approx \val{2.0}{\Msun}$ greatly constrains the EOS.  
On the one hand, the pressure near $\valrng[--]{2}{3}{\satdens}$ must be small enough to 
accommodate small radii;
on the other hand, the pressure must increase significantly at high densities in order 
to support massive neutron stars. 

Unfortunately, there are important complications in applying the Stefan-Boltzmann law 
[Eq.\,(\ref{e.quiescent-angular-area})] to extract 
neutron-star radii. First, in order to determine the luminosity $L_{\infty}$ one needs 
an accurate distance to the star. 
Whereas globular clusters provide reliable distance estimates, systematic differences 
at the $\approx\!20\%$ level 
remain between various techniques\,\cite[see Ref.][and references therein]
{Heyl2012Deep-Hubble-Spa}. Second, 
one must correct for interstellar absorption, which may be important in the far-
ultraviolet and soft X-ray energies.
The systematic uncertainty associated to the distance measurement should be eliminated 
with the imminent launch 
of the \emph{Gaia} spacecraft by the European Space Agency. \emph{Gaia} will determine 
parallax angles (i.e., trigonometric 
distances) with an unprecedented accuracy of 2\% (or better) for millions of stars on 
these globular 
clusters\,\citep[see Ref.][and references therein]{Pancino2013Globular-cluste}. 
However, relying on the Stefan-Boltzmann 
law assumes that neutron stars may be modeled as black bodies---which in general they 
are not. Indeed, one must fit the
observed spectrum to a model atmosphere that depends on the composition and perhaps on 
the stellar magnetic field. In 
addition, by using a single effective temperature $\Teffinf$ in Eq.\,(\ref{e.quiescent-angular-area}) one 
assumes spherical symmetry.   
If the temperature distribution is anisotropic, however, (for example from anisotropic heat 
conduction in a strong magnetic field) then the extracted stellar
radius may be inaccurate.  Note that there is no evidence for a
strong magnetic field and pulsations in the X-ray flux are not observed.

Perhaps the simplest way to accommodate neutron stars with small radii as in Eq.~(\ref{R_bob}) is for the EOS
to be soft up to densities of $\dens\lesssim\valrng[--]{2}{3}{\satdens}$. This 
would suggest that the neutron
skin-thickness of \lead[208] should also be small---a fact that can be directly tested 
via parity violating electron 
scattering at Jefferson Lab; see Sec.~\ref{subsec:neutronskins}. Alternatively---and 
more intriguing---the 
pressure at low densities could be large, leading to a thick neutron skin in 
\lead[208]. This would require a rapid 
softening of the EOS in order for the pressure at \valrng[--]{2}{3}{\satdens} to be 
small enough to accommodate 
Eq.\,(\ref{R_bob}). Such rapid density dependence could come from a change in the 
structure of dense matter, 
possibly from a phase transition, that could be probed with heavy-ion collisions. Note 
that regardless of the
softening mechanism, the EOS must stiffen again at higher densities in order to 
support massive neutron stars.

In summary, the following three observables are sensitive to the EOS at different 
densities. First, the neutron-skin 
thickness of \lead[208] is sensitive to the pressure at $\satdens$ and below. Second, 
the radius of an 
$\val{1.4}{\Msun}$ neutron star is sensitive to the pressure over a range of 
densities, but is most sensitive to 
densities in the range $\valrng[--]{2}{3}{\satdens}$ \citep{lattimer.prakash:neutron}.  
Finally the maximum mass of a 
neutron star is particularly sensitive to the pressure at high densities. Therefore, 
the density dependence of the EOS 
can be deduced by comparing these three observables.

\subsection{Neutron star cooling}
\label{subsec:NScooling}
Modeling of cooling neutron stars \citep{Tsuruta:1965py} predated their first detection.
For an exhaustive review of neutron star cooling, see Ref.~\citenum{yakovlev.pethick:neutron}; here we just give a summary of the salient features involving the nuclear symmetry energy.
The core of a neutron star is in $\beta$-equilibrium. To maintain this equilibrium, the following reactions are in balance:
\begin{eqnarray}
\nt &\to& \pt + \ele + \bar{\nu}_{e} \label{e.urca1}\\
\pt + \ele &\to& \nt + \nu_{e}. \label{e.urca2}
\end{eqnarray}
For degenerate $\nt\pt\ele$ matter, integration of the cross-section over the available phase space gives a characteristic $T^{6}$ temperature dependence to the neutrino emissivity, known as the direct Urca (dUrca) process. The reactions (\ref{e.urca1}) and (\ref{e.urca2}) cannot, however, simultaneously satisfy momentum and energy conservation unless the proton fraction $x\equiv \dens_{\mathrm{p}}/(\dens_{\mathrm{p}}+\dens_{\mathrm{n}})$ is greater than approximately $0.11$ \citep{lattimer91}.  As a result, for sufficiently low proton fraction the reaction proceeds via a bystander particle (for example),
\begin{eqnarray}
\nt +\nt 
	&\to& \nt  + \pt + \ele + \bar{\nu}_{e} \label{e.murca1}\\
\nt + \pt + \ele
	&\to& \nt + \nt + \nu_{e}, \label{e.murca2}
\end{eqnarray}
and similar reactions for which the bystander particle is a proton.
These reactions have a $T^{8}$ temperature dependance and have a much lower rate at temperatures $\lesssim \val{0.1}{\Giga\K}$, which are typical core temperatures for observed cooling neutron stars.

The pressure of nuclear matter at densities near saturation is proportional to $\partial S/\partial \dens$.
If the symmetry pressure increases rapidly with density, then it is energetically favorable for the proton fraction to be large. 
Moreover, if this trend holds at larger densities, then the interior of the neutron star may reach a proton fraction $x > 0.11$ before 
the star reaches its maximum mass.  
If so, then there is a neutron star mass threshold $\MdUrca$ above which direct Urca reactions [Eq.~(\ref{e.urca1})--(\ref{e.urca2})] can occur.
To date, all observed isolated cooling neutron stars are consistent with standard (i.e., no enhanced or dUrca) cooling \citep{Page2009Neutrino-Emissi}).   Note that enhanced cooling could make isolated neutron stars so cold and dim that they are not observed.  Indeed, there are many supernova remnants with no detected central compact object \citep{Kaplan2004An-X-Ray-Search,Kaplan2006An-X-Ray-Search}. While it is thought unlikely that all of the missing sources could be black holes, our knowledge of stellar evolution and collapse is insufficient to draw firmer conclusions. 

Quiescent neutron stars in transient low-mass X-ray binaries (qLMXBs) offer another test for the presence of enhanced cooling.  When the neutron star accretes, compression of matter in the crust raises the electron and neutron chemical potentials, which induce nuclear reactions \citep{sato79,haensel90a,Steiner2012Deep-crustal-he} that in turn heat the neutron star crust and core.  The crust and core temperatures come into equilibrium \citep{brown98:transients} in which the heat deposited in the crust is radiated via neutrinos from the core or photons from the surface during quiescent periods when the accretion rate is very low or zero. Observations of qLMXBs can therefore inform us about the interior temperature and the strength of neutrino emission in the neutron star \citep{yakovlev.levenfish.ea:coldest}.

Most observed qLMXBs are consistent with having standard cooling \citep{Heinke2007Constraints-on-}, given the large uncertainties in the long-term mean mass accretion rates.  However, there are two notable exceptions.  The first is SAX J1808.4$-$3658, which is a pulsing, accreting neutron star.  For this source only an upper limit on the thermal component of the luminosity is reported, at \val{\sci{1.1}{31}}{\ergspersecond} \citep{Heinke2007Constraints-on-}.  This implies a core temperature \valrng[--]{20}{30}{\,\Mega\K} \citep{brown.bildsten.ea:variability}.  If this core temperature is set by balancing accretion-induced heating with neutrino cooling then the presence of enhanced cooling is required.  An even colder neutron star is in the transient 1H~1905$+$00 for which only an upper limit of \val{\sci{2.4}{30}}{\ergspersecond} on the luminosity is reported \citep{Jonker2007The-Cold-Neutro}.  For this system, the time-averaged mass accretion rate is unknown, unfortunately, so more quantitative evaluations of the neutrino cooling are not possible.

If enhanced cooling is not observed, this suggests that the symmetry energy at high densities is not large so that dUrca is not possible.  Furthermore, the beta decays of additional hadrons such as hyperons, if present, are suppressed (for example by large pairing gaps).   Alternatively, if enhanced cooling is confirmed, in at least some neutron stars, then one can constrain the hadrons responsible for the cooling.  Determining the high density symmetry energy from heavy ion collisions, Sec. \ref{sec:HI}, can rule in or rule out dUrca.  If dUrca is ruled out, this suggests the enhanced cooling is from the beta decay of other hadrons such as hyperons, mesons, or quarks.  This would demonstrate that dense matter contains hadrons in addition to just neutrons and protons.

\section{The symmetry energy and heavy-ion collisions}
\label{sec:HI}

Heavy ion collisions provide the only means to investigate the behavior of the symmetry energy within a laboratory environment by 
compressing nuclear matter from moderate to high densities. At low incident energies---from 30 to 150 MeV per nucleon---expansion 
after the initial compression creates a low density region where symmetry-energy effects including cluster formation can be studied. The 
forces resulting from such a compression and the subsequent expansion influence strongly the motion of the ejected matter and provide 
observables that are sensitive to the EOS. Extracting information of relevance to the symmetry energy presents a serious challenge 
given the relatively low neutron-proton asymmetry of the colliding ions. 
To overcome this obstacle one makes use of the fact that the forces associated with the symmetry energy affect differently
the various members of an isospin multiplet. Enhanced sensitivity to the density dependence of the symmetry energy is then 
obtained by comparing neutron- against proton-like observables. As full equilibrium is often not reached in heavy-ion collisions,  
experimental analyses depend heavily on theoretical transport models.

The dynamical theories underlying these transport models rely on various assumptions and approximations. For the 
most part these models are based either on a semi-classical approximation to the time-dependent Hartree-Fock equations
[Boltzmann-Uehling-Uhlenbeck (BUU)] or on Molecular Dynamics simulations that incorporate quantum effects [such as 
Quantum Molecular Dynamics (QMD) and Antisymmetrized Molecular Dynamics (AMD)]. These models
include a mean-field potential usually with momentum dependence plus a collision term with in-medium nucleon-nucleon cross sections, which describes the dissipative features of the 
collision.  The mean field also includes a ``tunable'' density dependence for the symmetry energy. 
Since the transport equations are not solvable directly, they are treated by simulations. 
 Given their critical role in the interpretation of heavy ion collisions, we provide in a separate subsection 
(Sec.~\ref{sec:HI_transp}) a brief description of transport models, their current development, and ways 
to improve the consistency and reliability of their predictions.

Investigations on the EOS of {\it symmetric} nuclear matter have been very successful in understanding
 a wide set
of observables---such as flows of nucleons and light particles, and kaon 
production\,\cite{danielewicz02,fuchs06}---from transport-model predictions. These efforts have provided robust constraints 
on the EOS of symmetric nuclear matter. A similar comprehensive effort is now needed to place more stringent constraints 
on the density dependence of the symmetry energy. Consequently heavy-ion collisions have increasingly been used in recent years
to probe the symmetry energy over a wide range of densities by varying the incident energy, impact parameter, and
isospin asymmetry of the colliding system. In the following sections observables used to probe the symmetry 
energy are introduced, first at sub-saturation densities, followed by a discussion of the observables proposed 
for super-saturation densities. In each density domain, difficulties in the theoretical 
description of these observables are identified and strategies to clarify them and improve their predictive power are 
discussed.

\subsection{Heavy-ion collisions at low densities}
\label{sec:HI_lowdens}

Heavy-ion collisions 
at incident energies from about 35 MeV to 150 MeV per nucleon give access to the symmetry energy at densities from about 50\% above $\satdens$ down to about $0.1\satdens$. The initially compressed nuclei expand towards low densities where 
many intermediate mass fragments are formed. Nucleons and light clusters such as deuterons, tritons, and alpha particles
are emitted during the fragmentation process as well as from the excited primary fragments. By selecting collision 
geometries from peripheral to central collisions, one can study a number of phenomena which depend on the transport 
of isospin and thus provide information on the density dependence of the symmetry energy.

In semi-peripheral collisions the amount of isospin diffusion ($N/Z$ equilibration) between collision partners of different 
neutron-proton asymmetries is driven by the symmetry energy at subsaturation densities. The degree of isospin diffusion is quantified 
in terms of an isospin transport ratio by comparing reaction systems with different combinations of neutron-rich and 
neutron-poor projectiles and targets. For example, by using heavy-ion collision data of $\tin[112]$ and $\tin[124]$ in identical 
and mixed combinations at 50 MeV/nucleon, constraints on the symmetry energy were obtained for densities in the range
$\dens=\satdens/3$ to $\satdens$ using several isospin diffusion observable \cite{tsang:2009}.

\begin{figure*}[htbp]
\centering
\includegraphics[width=\figwidth]{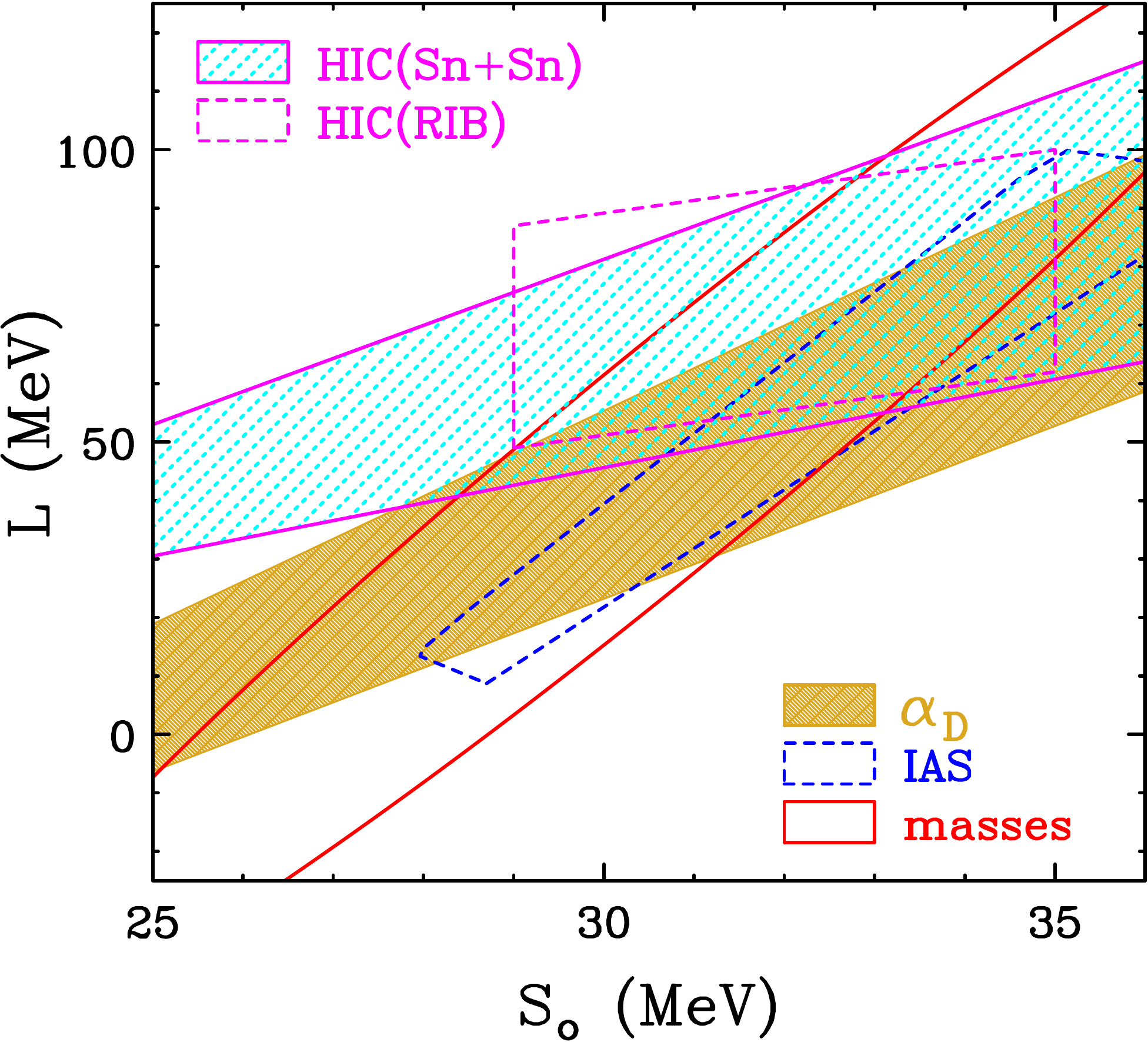}\hspace{0.05\textwidth}
\includegraphics[width=\figwidth]{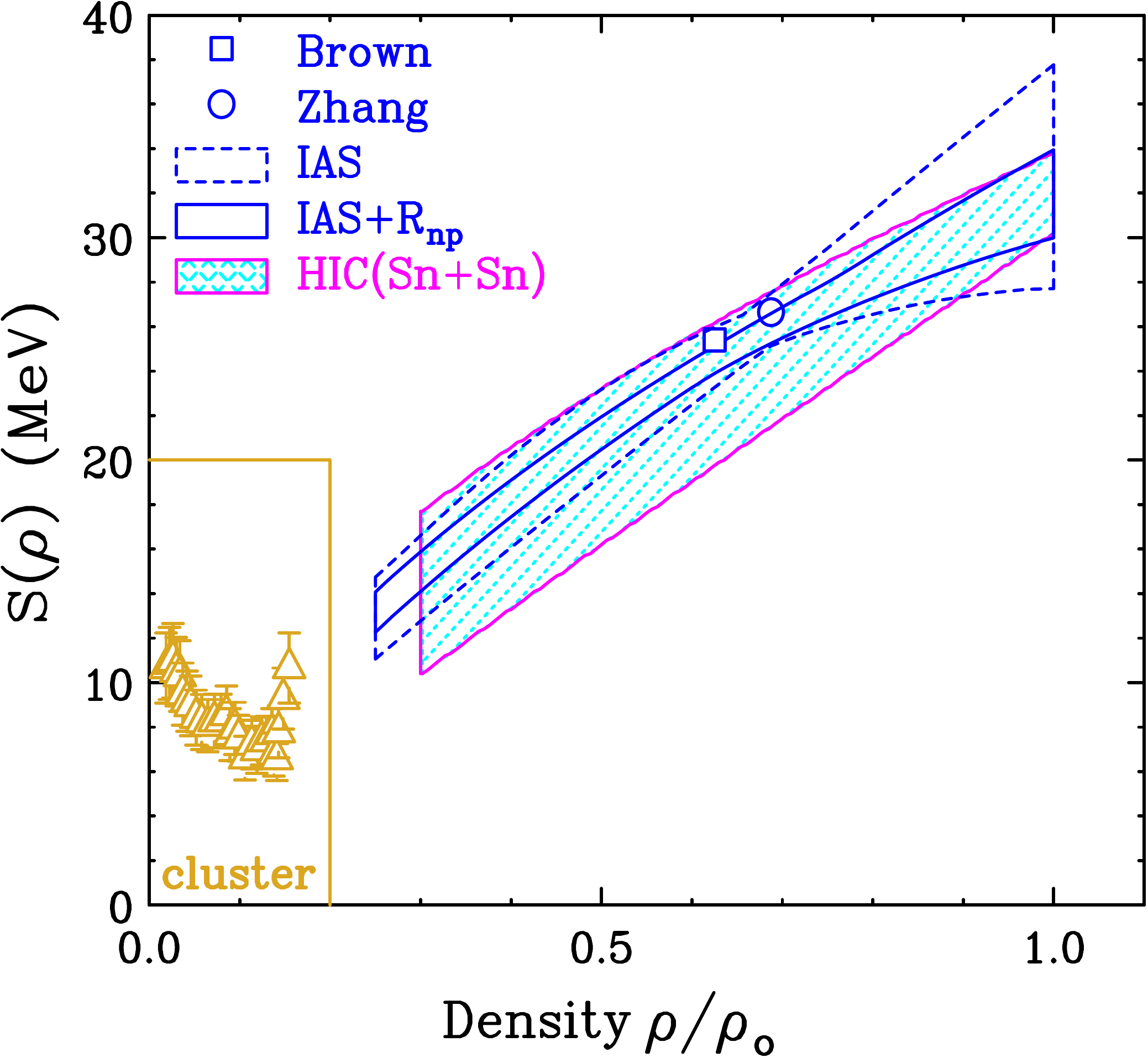}
\caption{Constraints on the density dependence of the symmetry energy from both heavy-ion collisions
and nuclear-structure observables.
\emph{Left:} 
Correlation between $S_{\!0}$ and the slope of the symmetry energy $L$ at saturation density (see text for 
a detailed description).
\emph{Right:} The symmetry energy $S(\dens)$ as a function of baryon density (see text for a detailed description).}
\label{constraints}
\end{figure*}

Constraints on the density dependence of the symmetry energy obtained from heavy-ion experiments are shown 
in Fig.~\ref{constraints} in two representations 
and are compared against those obtained from nuclear-structure observables discussed in Sec.\,\ref{sec:nuclear_structure}.
On the left-hand panel of Fig.\,\ref{constraints} we display constraints on the symmetry energy $S_{\!0}\!\equiv\!J$ and 
its slope $L$ at saturation density.  
The blue hatched area labelled HIC(Sn+Sn) was determined from 
isospin diffusion observables measured in mid-peripheral collisions of Sn isotopes\,\cite{tsang:2009}. A constraint on the
symmetry energy obtained in recent measurements of the mean $N/Z$ distributions of the emitted fragments with radioactive 
ion beams of $\magnesium[32]$ on a $\beryllium[9]$ target at \val{73}{\MeV} per nucleon is shown by the area enclosed by 
the dashed purple line labelled HIC(RIB)\,\cite{kohley13}. (Note, that the limits of $S_0$ in these two areas only indicate the range of values used in the transport simulations and are not to be interpreted as 
experimental limits on $S_0$ from HICs.) 
These constraints from HICs are compared against those obtained from nuclear structure;
in particular, from studies of (a) isobaric analog resonances (blue dashed polygon) \citep{Danielewicz13}, (b) the electric dipole 
polarizability in $\lead[208]$ (gold shaded region) \cite{Tamii:2011pv,Roca-Maza:2013mla,tamii13} both with better than $90\%$ confidence limit (Cl), and (c) nuclear binding 
energies using the UNEDF0 energy density functional (two red curves forming part of an ellipsoid, about 90\% Cl) \cite{Kortelainen:2010hv}. 
In all cases we observe a strong correlation between $S_{\!0}$ and $L$. This suggests that the accurate determination of 
one will place stringent constraints on the other. 

 The strong correlation between $S_{\!0}$ and $L$ has a simple origin: 
It has been recognized for a long time that nuclear ground-state properties---especially nuclear masses of 
neutron-rich nuclei---determine rather accurately the value of the symmetry energy at a sub-saturation of
$\dens\!\approx\!\val{0.1}{\fermi^{-3}}\!\approx\!(2/3)\satdens$.
In particular, Ref.~\citep{Brown2013Constraints-on-} using the family of Skyrme CSkp functionals to fit ground-state 
properties of double magic nuclei found a value of $S(\dens\!=\!\val{0.1}{\fermi^{-3}})\!=\val{25.4\pm0.8}{\MeV}$; 
this value is plotted as an open square in the right panel of Fig.~\ref{constraints}. A comparable analysis by \citep{Zhang2013Constraining-th} 
using the masses of 38 spherical nuclei gives a value of the symmetry energy at a slightly larger density of 
$S(\dens\!=\!\val{0.11}{\fermi^{-3}})\!=\val{26.65\pm0.2}{\MeV}$ (plotted as an open circle). 
By invoking Eq.\,(\ref{JLK}) one then obtains
\begin{equation}
 S(\dens\!=\!\val{0.1}{\fermi^{-3}}) = S_{\!0} - \frac{L}{9} + \frac{K_{\rm sym}}{162} + \ldots
 \approx  S_{\!0} - \frac{L}{9} 
\label{JL}
\end{equation}
Thus the accurate determination of 
$S(\dens\!=\!\val{0.1}{\fermi^{-3}})$ from the ground-state properties of finite nuclei leads to the strong correlation
between  $S_0$ and $L$.
The search for values of $S_0$ and $L$ is therefore rather a one-dimensional problem along this correlation.

On the right-hand panel of Fig.\,\ref{constraints} we show a different interpretation of the constraints by focusing 
directly on the density dependence of the symmetry energy $S(\dens)$.
The shaded area labelled HIC(Sn+Sn) results from the analysis also shown on the left of isospin diffusion observables from
Ref.\,\cite{tsang:2009}.
From the analysis of isobaric analog states (IAS) by\,\citep{Danielewicz13} two constraints have been reported.
The area enclosed by the dashed blue line comes from the same IAS analysis shown on the left-hand panel. 
The area enclosed by the solid blue line results when the IAS analysis is supplemented with additional
constrains from neutron-skin data $R_{\rm skin}$. Note that the addition of neutron-skin information reduces 
significantly the constraint area---especially near saturation density. This is natural given the strong
correlation between $R_{\rm skin}$ and the slope of the symmetry energy $L$, noted earlier. However, one should 
remember that
the extraction of $R_{\rm skin}$ from hadronic experiments is beset by large theoretical uncertainties.

The symmetry energy at very low density obtained from the analysis of cluster production in HICs as discussed in 
Sec.\,\ref{sec:verylowdensity} is shown enclosed by a box in the right-hand panel of Fig.\,\ref{constraints}.  Here it should be 
noted that this analysis determines the global symmetry energy including cluster correlations in non-uniform matter at a 
temperature of several MeV, depending on the density. Therefore it cannot be compared directly to the 
symmetry energy of uniform matter at zero temperature, which is shown at the higher densities.

Alltogether we see from Fig.\,\ref{constraints}  that heavy-ion results are consistent with those extracted from nuclear-structure information. This lends credibility 
to the use of heavy-ion collisions as a probe of the symmetry energy at supra-saturation densities, a region inaccessible to 
nuclear-structure experiments (see Sec.\,\ref{sec:HI_hidens}).

In central collisions of neutron-rich systems one observes isospin fractionation, namely, a differentiation of the isospin 
distribution among the reaction products\,\cite{colonna02,tsang03}. In a neutron-rich environment the neutron mean-field
potential is more repulsive than that of the proton. This enhances the neutron emission in such a way that the 
neutron-to-proton ratio of the emitted particles (gas) is larger than that of the formed fragments (liquid). Then, the spectra 
and the spectral ratios of emitted particles should be a direct probe of the symmetry energy. 
The experimental ratios  at low
energies in the center-of-mass system are large, much beyond Coulomb effect
expectations. 
However, a complete 
understanding of this phenomenon has not yet been achieved because of two aspects of the transport models. The
first one is associated with the isovector momentum dependence---also expressed as the neutron-proton effective mass
difference---which is highly uncertain and influences the high-energy part of the particle spectra. The second one
involves the treatment of cluster formation which affects the low-energy particle spectra. As discussed in 
Sec.\,\ref{sec:verylowdensity} and in Ref.\,\cite{typel10}, cluster formation increases the symmetry energy at very low 
densities and may affect neutrino transport in the neutrinosphere.
Cluster formation is important in heavy-ion collisions---especially in the expansion phase. For example, in the final 
state of $\mathrm{Sn\!+\!Sn}$ collisions at 50 MeV/nucleon, about 90\% of the protons in the system are bound in clusters 
and heavier fragments. However, the treatment of cluster formation in transport theories, 
including how to take into account the physical 
level density of the in-medium clusters which may contain bound and resonance states, is a difficult problem, which will be
discussed more in Sec.\,\ref{sec:HI_transp}.

Neutron detection efficiencies are often uncertain, thus ratios of {\it n/p} ratios, i.e. double ratios, of systems with different asymmetry have been obtained, which are hopefully less 
affected by the neutron efficiency, but some of the sensitivity of the single  {\it n/p} ratio may be 
reduced in the double ratios. Alternatively, one has measured $t/^3He$ ratios, which display similar trends as the {\it n/p} ratios. 
However, their theoretical description requires a better description of the cluster production mechanism.
For example, a recent theoretical and experimental study finds that the formation of alphas is important in describing  the 
triton to \helium[3] spectra and spectrum ratio. 
The problem of describing cluster formation  can be somewhat alleviated by constructing
``coalescence invariant'' quantities, i.e., observables summed over all light clusters, which show much better agreement between theory and experiment \cite{Coupland:2011,zhang12}.

Furthermore intermediate mass fragments are copiously emitted in lower energy heavy ion collisions, 
which is a process not only interesting 
in itself but which also contains information on the symmetry energy. It has been shown \cite{Amorini:2009,Cardella:2012}
that different reaction mechanisms from isospin fractionation in the disintegrating system to isospin migration to the 
low-density neck region , as well as cluster correations, depend on the density-dependence of the symmetry energy 
\cite{Coupland:2011,Rizzo:2008,DeFilippo:2012}.   

In the current situation, possible ways to go forward are (1)
to find observables which can be predicted with minimum uncertainty from cluster production by
available transport models, (2) to compare the predictions
of different models in detail until the physical origins of different
predictions become clear (see subsection on transport models, Sec.~\ref{sec:HI_transp}),
 (3) to establish improved models that
can consistently describe the global reaction dynamics (including
cluster and fragment formation) together with the observables sensitive to the
symmetry energy. Efforts for solutions
to the options (2) and (3) are particularly important.  Specifically, models should be able to describe many experimental observables simultaneously.

On the experimental front, measurements with larger isospin asymmetry made possible with the availability of high intensity rare isotope beams provide higher sensitivities to the symmetry energy and will be able to further clarify these issues and establish the understanding of particle emissions in neutron-rich systems.
An experiment that measures isospin diffusion using radioactive beams of \indium[109] on a \tin[124] target ($\Delta(\delta)=0.093$) has recently been performed at RIKEN. A more desirable reaction is to use a radioactive \tin[132] beam  on a \tin[112] target ($\Delta(\delta)=0.135$). By increasing the asymmetry by more than 50\%, improved uncertainty limits of the constraints are expected.

\subsection{Heavy-ion collisions at higher densities}
\label{sec:HI_hidens}
Although there has been enormous progress in elucidating the nature of cold hadronic matter at very high densities directly from 
 QCD\,\cite{Alford:1998mk}, the density region where these predictions apply is out of reach of experimental tests in
high energy heavy ion collisons and even at the enormous densities that exist in 
the core of neutron stars. Hence, for these density regions one must rely on theoretical models that, unfortunately, 
differ dramatically in their prediction of 
 the high-density behavior of the EOS\,\cite{Brown:2000}.
A major source of these difficulties is the incomplete knowledge of the 
short range isovector correlations \cite{BALi10}. 
In order to constrain the models one must rely on both laboratory experiments with energetic heavy ions and astronomical 
observations of neutron stars.

Observations of massive neutron 
stars\,\cite{Demorest2010A-two-solar-mas,Antoniadis2013A-Massive-Pulsa} imply that the high-density component of the EOS must 
be stiff. The recent analysis by \citep{Guillot2013Measurement-of-} seems to suggest that neutron stars have small radii,
implying a rather soft symmetry energy in the region of $\dens\!=\!\valrng[--]{2}{3}{\satdens}$ (see Sec.~\ref{sec:high_density}).
In combination, both of these results---large stellar masses and small radii---pose serious challenges to theories of hadronic 
matter without exotic degrees of freedom.

\begin{figure}[htbp]
\includegraphics[width=\figwidth]{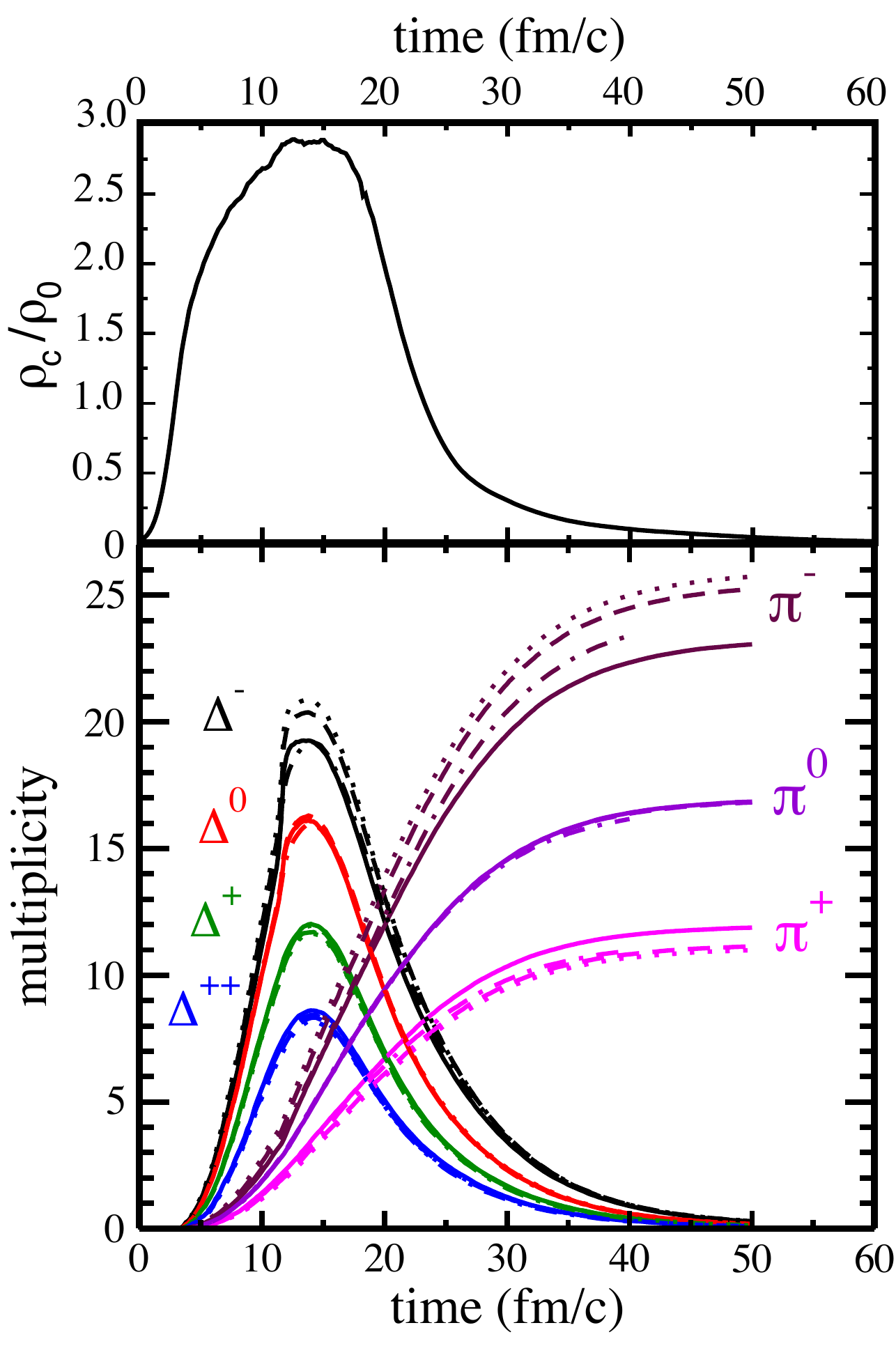}
\caption{Time evolution of a central $\mathrm{Au+Au}$ collision at en energy of $E\!=\!\val{1}{\GeV/\amu}$. 
Top panel: density in the central cell in units of saturation density $\satdens$.
Bottom panel: multiplicities of produced $\Delta^{(-,0,+,++)}$ and $\pi^{(-,0,+)}$ particles. The curves correspond 
to calculations with different assumptions on the symmetry energy: stiff (dotted), linear (dashed), soft (dot-dashed), 
only kinetic symmetry energy (solid)\,\cite{ferini06}.}
\label{evolution}
\end{figure}

Intermediate and high-energy (relativistic) heavy-ion collisions with rare isotope beams can be used to study neutron-rich systems 
that attain this range of densities for short-time intervals and may help elucidate the nature of these puzzles.
At collision energies from about a hundred MeV to a few GeV, nuclear matter is compressed from about 1.5 up to $3.5 \satdens$ in 
a short time interval of about \valrng[--]{10}{100}{\fermi/c}  or about $\val{10^{-22}}{\second}$. For example, the top panel of Fig.\,\ref{evolution} shows 
the time evolution of the central density in a $\mathrm{Au+Au}$ collision at \val{1}{\GeV/\amu}, suggesting that nucleonic matter may
be compressed to almost $3\satdens$ for a period of about \valrng[--]{5}{10}{\fermi/c}.

The big challenge to theoretical descriptions of heavy-ion collisions is the extraction of properties relevant
to the high-density phase from the experimental observables constructed from the particles detected in the final stages 
of the reactions. As in the case of low-density matter, the aim is to magnify the impact of the symmetry energy by measuring 
differences or ratios of observables involving charge-symmetric pairs of particles, such as $ n/p, t/\helium[3]$ or $\pi^{-}/\pi^{+}$.
In anticipation of these critical experiments, 
at least two time projection chambers (TPC) are being commissioned: the Active Target Time Projection Chamber (AT-TPC) at NSCL/FRIB and the SAMURAI-TPC detector at RIKEN, where an experimental program involving central collisions of $\tin[132]+\tin[124]$ and $\tin[124]+\tin[112]$ at 200 and 300 MeV per nucleon is being planned.

There already exist some data and corresponding transport-model predictions for flow and pion ratio observables, but the available 
experimental and theoretical information is far from complete. The existing data (from the Kaos and FOPI collaborations at GSI) 
involve nucleon, light cluster, and pion observables measured mainly in $\mathrm{Au\!+\!Au}$ reactions at energies between 400 MeV 
and 1.5 GeV per nucleon\,\cite{Reisdorf-FOPI-2007,Reisdorf-FOPI-2012}. These data were primarily obtained and analyzed for investigating the EOS of symmetric matter. However, the scarcity of neutron 
data and the lack of projectile-target systems with a wide range of neutron-proton asymmetries limits the sensitivity of these data 
to the symmetry energy. New measurements have been and will continue to be performed to overcome these limitations. Beams of 
widely different asymmetries from advanced rare isotope facilities will play a pivotal role in future studies by providing selective 
sensitivity to the symmetry energy while minimizing variations in the contributions from the symmetric matter EOS. In addition to 
studying the collisions of Sn isotopes including unstable neutron rich (e.g., \tin[132]) and neutron poor (e.g., \tin[108]) beams, a 
program is being developed to measure reaction systems with the same mass, such as  $\tin[112]\!+\!\tin[112]$ and 
$\ruthenium[112]\!+\!\tin[112]$ and highly asymmetric systems such as $\calcium[36]\!+\!\tin[112]$ and $\calcium[52]\!+\!\tin[124]$. 
These very asymmetric reactions provide a larger lever arm in the determination of critical isospin-dependent inputs for transport 
models, such as the symmetry potentials and the in-medium nucleon-nucleon collisions cross sections.

Similarly as at lower densities, primary discriminators of symmetry-energy effects are expected to involve differences 
in neutron and proton observables,
such as spectra and spectral ratios of emitted neutrons and protons or $t$ and \helium[3], as well as flow differences. 
The differences in neutron and proton potentials including the momentum dependence or the effective mass difference
influence the asymmetry of the flow and the production of secondary particles.
Some calculations suggest that at higher densities ratios of isospin yields and flows may be more sensitive to the nucleon 
mass splitting than to the density dependence of the symmetry energy\,\cite{colonna10}. 
If so, this provides an opportunity to determine separately both the momentum dependence of the interaction 
and the density dependence of the symmetry energy at high densities. 

The collective motion induced by the pressure developed in the compressed region is quantified in terms of a Fourier series 
of the azimuthal distribution, where the first two coefficients are known as the transverse and elliptic flow. Of these, the elliptic 
flow probes the emission perpendicular to the reaction plane, and is therefore sensitive to pressure gradients in 
the high-density region of the collision. Thus, differences (or ratios) of the neutron-proton elliptic flow are regarded as sensitive
probes of the high-density component of the symmetry energy. Indeed, there are indications that the elliptic-flow ratio of
neutrons to protons is sensitive to the symmetry energy at supra-normal densities from the analysis of FOPI/LAND 
experiments at GSI\,\cite{russotto11,Cozma:2013}.

In high-energy collisions new particles are copiously produced. In particular, the ratio of negatively to positively charged 
pions $\pi^-/\pi^+$ has been proposed as a probe of the symmetry energy at high densities \cite{ferini05,xiao09}.
Two estimates of pion production---the isobar model and the assumption of chemical equilibrium---both predict 
a strong sensitivity to the asymmetry of the medium and thus to the symmetry energy.
In heavy-ion collisions pions are primarily produced in NN-collisions via the excitation and subsequent decay of the 
$\Delta$-resonance. 
If the medium is neutron rich, the more abundant $nn$ collisions will increase the production of $\Delta^{0}$ and $\Delta^{-}$ 
resonances, leading ultimately to an enhanced $\pi^-/\pi^+$ ratio. 
For example, the excitation of the $\Delta^{-}$ resonance and its ensuing 
$n\pi^{-}$ decay is driven exclusively by the $nn\rightarrow p\Delta^{-}$ reaction and thus should significantly enhance the 
$\pi^-/\pi^+$ ratio in a neutron-rich medium.
Some theoretical analyses moreover predict that the mean field effects and the threshold effects in 
the production of the particles influence the ratio in opposite directions, making predictions 
more critical \cite{ferini05}.  
Thus, imposing meaningful constraints on the symmetry 
energy depends crucially on a good understanding of the $\pi\textrm{-}\Delta$ dynamics in the medium. This constitutes a serious theoretical
challenge, as, e.g., one must understand how the pion and $\Delta$ propagators (mass and width) are modified in the 
nuclear medium. Moreover, the 
propagation of particles with finite width has to be approximated. It has been noted that the sensitivity of pion production to the 
symmetry energy increases at lower incident energies where the in-medium masses of nucleons, Deltas, and pions 
(i.e. the real parts of their self-energies) provide critical threshold effects for particle production. In the bottom panel of 
Fig.\,\ref{evolution} we display the time evolution of both the $\Delta$ and pion multiplicities. Comparing to the upper panel
it is seen that Deltas exist mainly during the dense phase of the collision, while the pion multiplicities change also during 
the expansion phase due to reabsorption and charge exchange scattering. 

At present it seems that the different transport codes make different assumptions about the relevant self-energies for nucleons, Deltas, and 
pions as well as for the elastic and inelastic cross sections. This may contribute to the large differences in the model predictions 
of the pion ratio\,\cite{xiao09,feng10,ferini05,jhong13}. Currently, it is not clear whether a stiff or a soft symmetry energy 
provides a better description of the data. Indeed, in the calculations of Ref.~\citep{xiao09} a very soft symmetry energy was 
postulated for the explanation of the experimental data, in strong disagreement with the conclusions from flow measurements \cite{russotto11} . 
Clearly, the current situation needs to be clarified. There is also a need for more data on pion 
spectra and pion flow for systems with fixed total charge and different asymmetry---to clearly distinguish between 
Coulomb and symmetry-energy effects. 
Such experiments will provide valuable constraints on transport models and should help to understand the situation. 
Eventually, pion yields should become a very useful source of information on the symmetry energy.

Pions are created and reabsorbed via the formation of $\Delta$ resonances. Reabsorption and pionic charge exchange increase 
with energy, and thus decrease the sensitivity to the high density phase of the collision, 
as seen in the bottom panel of Fig.~\ref{evolution}. In this 
respect the investigation of kaon production could be advantageous and has been proposed as a sensitive probe of the symmetry 
energy\,\cite{ferini05}.  
In contrast to pions, $K^{0}$ and $K^{+}$ mesons---which carry a strange antiquark---interact weakly with matter and have
rather long mean-free paths and should therefore be good probes of the high density phase of the collision. In fact, the ratio of $K^+$ mesons produced in collisions of light to heavy nuclei has provided in the past a 
robust observable for the stiffness of the EOS for symmetric matter\,\cite{fuchs06}. Similarly, the ratio of $K^{0}$ (produced 
mainly from $nn$ collisions) to $K^{+}$ (produced mainly from $pp$ collisions) has been predicted to be sensitive  to 
the high-density behavior of the symmetry energy \cite{ferini06}. As in the case of pion production, the strongest sensitivity is expected near 
the kaon-production threshold, as many kaons are produced via pion-baryon reactions in the nuclear medium. To isolate 
symmetry-energy effects, the density and isospin dependence of the kaon self-energy must be properly understood\,\cite{gaitanos10}. 
Past measurements\,\cite{lopez07} suffered from very different acceptances for $K^0$ and $K^+$, requiring double ratios, but  
high-statistics measurements may now be possible with the HADES detector at GSI.

\subsection{Transport Codes }
\label{sec:HI_transp}

Transport models follow the evolution of the colliding system under the influence of mean-field potentials and of dissipation
(encoded in a collision term). These models have been very successful for many years in both predicting and explaining many 
phenomena in heavy-ion reactions. However, in spite of this success questions remain---partly related to the formulation 
and partly to the implementation of the models. Given the critical role that transport models have in interpreting the experimental
results of heavy-ion collisions, a way forward in the study of the
symmetry energy requires an increase in the predictive power and reliability of transport 
codes. In this subsection we address some of these difficulties and propose ways to overcome them.

Different transport approaches have been developed over the years. Boltzmann-like approaches, known, e.g., under the names of Boltzmann-Uehling-Uhlenbeck (BUU) 
or Stochastic Mean Field (SMF), describe the evolution of the single particle phase space 
density. In principle, they have to be augmented by a fluctuation term, leading to the Boltzmann-Langevin equation; various 
attempts in this directions have been put forward \citep[see Ref.][and references therein]{Colonna2013Fluctuations-an}. 
On the other hand, 
Molecular Dynamics approaches start from classical molecular dynamics and introduce finite size wave-packets, 
usually without anti-symmetrization (Quantum Molecular Dynamics, QMD). Collisions between these wave-packets lead to dissipation and to fluctuations
which are controlled by the width of the particles. An significant further development is Anti-symmetrized Molecular Dynamics (AMD) 
[and approximately 
Constrained Molecular Dynamics (CoMD)], which includes anti-symmetrization between the wave packets, and thus represents a 
quantum-mechanical transport model. The essential difference between Boltzmann and molecular dynamics approaches is the amount of 
fluctuation introduced into the evolution of the system. In general, fluctuations are important when the evolution enters a spinodal region of 
the phase diagram, where the phase transition leads to the formation of fragments or light clusters. This is very important 
in collisions in the low density region (see subsection\,\ref{sec:HI_lowdens}) but also at higher densities clusters constitute an 
essential fraction of the ejectiles. A reliable description of cluster production is therefore crucial in the comparison with experiments. 

The effect of cluster formation on the reaction dynamics and of isospin migration driven by both density and isospin 
gradients has been studied using a version of the BUU approach that treats clusters (deuterons, tritons, and \helium[3], 
but not alpha particles) as distinguishable particles\,\cite{Coupland:2011}. The collision terms couple the equations for the
different particles. In the QMD and AMD approaches which employ nucleon wave
packets, a cluster can be naturally described by placing
the appropriate wave packets at the same phase space point.
However, the
probability of forming a cluster is then governed by the classical phase
space in which the quantum bound state or resonance contribution is
missing.  In a recent version of AMD, the collision
term has been improved in order to better incorporate the
probability that one (or both) of the colliding nucleons form a
cluster (with $A = 2$, 3 and 4) with other particles in the system
\cite{ono13-1,ono13-2}.
Both BUU and AMD calculations including clusters
demonstrate that cluster formation can change the reaction
dynamics and fragmentation mechanism \cite{Coupland:2011,ono13-1,ono13-2}. This effect may explain the
differences in the isospin transport ratios obtained with usual BUU calculations \citep{Li:2005jy} 
relative to QMD calculations which have different cluster formation probabilities \cite{tsang:2009}.

The dynamical evolution of particles with finite width (``off-shell transport'') is in principle understood in the framework of 
the Kadanoff-Baym equations, but usually neglected or treated in simple approximations. It is expected, however, to be important 
in the sub- or near-threshold production of particles, like pions, $\Delta$-resonances, or strange particles, which are important 
probes of the symmetry energy at higher density.

The physical input into transport simulations are the mean field potential, often derived from an energy density functional, and the 
in-medium elastic and inelastic cross sections. Both of these are often parametrized independently of each other, even though, in 
principle, they are related by a consistent approximation to the in-medium effective $T$-matrix, e.g., in the Brueckner-Hartree-Fock 
approximation. A consistent approach in all models is a highly desirable goal.
We note that an important check involves solving the transport models in a confined box to simulate the conditions of
thermally equilibrated systems. This can test the relation between the adopted effective interaction and the corresponding 
EOS \cite{ono06,baran98,papa13}. Moreover, it can provide useful links to dynamical conditions, such as those
found in the core and crust of neutrons stars and in supernovae explosions.

In all approaches, the set of transport equations constitutes a complex system of non-linear integro-differential equations which 
cannot be solved in closed form. One usually relies on simulations, for example, by test-particle methods in the BUU approach. 
In particular, the collision term is evaluated stochastically. In such simulations, some of the implementations are not necessarily dictated by the 
underlying equations and are handled differently in the numerical codes. Hence---even within the same theoretical framework--- 
different versions of codes with varying procedures and inputs have been developed, though are often not properly documented. 
Thus, seemingly similar calculations have led to substantially different results. Clearly, this is an unsatisfactory state of affairs. 
This situation has stimulated various efforts for code comparison
\cite{Kolomeitsev2005Transport-theor,Kohley:2012,Colonna:2010,Rizzo:2007}.
Earlier attempts to analyze the situation took place in workshops at Trento in 2004 (for particle production at 
\val{1}{\GeV/\amu}\,\citep{Kolomeitsev2005Transport-theor}) and in 2009 (for flow at \val{100}{\MeV/\amu} and 
\val{400}{\MeV/\amu}). During the 2009 workshop calculations of the main transport codes with identical physical 
inputs were compared with respect to the main global observables---such as yields, rapidity distributions and flow observables.
Although in general good qualitative agreement was found, the degree
of quantitative agreement requires improved understanding of the
calculated observables in order to test 
finer details of the EOS, particularly those related to the symmetry energy. To clarify the source of these discrepancies,  
internal quantities, such as the number, energy distribution, and blocking of collisions, were investigated and, indeed, showed 
large differences. This is significant, since differences in the collision term can strongly influence particle production near thresholds.

A consensus among the practitioners of the transport codes is that a continuation and stabilization of such comparisons is 
a worthy and timely project. Indeed, a new workshop in China where many groups are engaged in 
transport calculations using a variety of codes has been organized. The aim of the workshop is to compare, verify, and validate the reliability of 
the most widely used codes,
and to document properly 
versions of the codes together with benchmark examples. This endeavor should strengthen considerably
the impact of heavy-ion research on the investigation of the nuclear EOS---particularly on the symmetry energy at high 
densities. From the remaining differences between codes it will allow to estimate systematic theoretical uncertainities.

\section{Conclusions and a way forward}
\label{sec:conclusions}

The last few years have seen enormous progress in our understanding of the density
dependence of the symmetry energy. However, significant challenges lie ahead. The
present document provides a roadmap that continues to foster dialogue and promotes
collaborations between the astrophysics and the nuclear-physics communities. We
articulate a way forward in areas of relevance to the symmetry energy---such as nuclear
structure, heavy-ion collisions, and neutron stars---by proposing new terrestrial experiments
and astrophysical observations at next generation facilities. In what follows we summarize
our vision for the future of the symmetry energy.

\subsection{The way forward at very low densities}
The symmetry energy at very low densities is of great relevance to the neutrinosphere
region in core collapse supernovae. The neutrinosphere---the surface of last scattering
for the neutrinos---is composed of a low density gas of neutron-rich matter at temperatures
of about $T =\val{5}{\MeV}$ and densities of $\rho \approx 0.01\rhoz$. Under these
conditions neutron-rich matter is unstable against cluster formation, as the formation
of light nuclei is energetically favorable. The existence of light clusters could modify
the neutrino opacity and ultimately affect the conditions for nucleosynthesis. Remarkably,
many properties of the neutrinosphere may be simulated in the collision of heavy ions. New
experiments involving heavy ion collisions with very neutron-rich (or proton-rich) nuclei
at present and near future radioactive beam facilities will be able to reproduce many neutrinosphere conditions. Moreover, measurement of the yields and distribution of light
fragments may help elucidate the dynamics of the neutrinosphere. In particular,
comparing these yields to transport models with improved descriptions of light clusters
should allow more accurate supernova simulations of neutrino spectra and nucleosynthesis.


\subsection{The way forward with neutron skins}
The neutron-skin thickness of heavy nuclei is a strong isovector indicator that correlates
strongly to the density dependence of the symmetry energy---particularly to the symmetry
slope $L$. The sensitivity of  $R_{\rm skin}$ to $L$ emerges from a competition between
surface tension and the \emph{difference} between the symmetry energy at the center and
surface of the nucleus. In particular, for a stiff symmetry energy, it is favorable to push
the excess neutrons to the surface where the symmetry energy is small. Thus,
models with large $L$ tend to produce thicker skins. The upcoming PREX-II
experiment promises to determine the neutron-skin thickness of \lead[208]
with a $\pm\val{0.06}{\fermi}$ accuracy. Using the strong correlation between
$\RskinPb$ and $L$, as evinced in Fig.~\ref{FigNSkins}, leads to a
determination of $L$ with an accuracy of $\Delta L_{\rm exp}=\val{40.8}{\MeV}$.
Given that the model predictions suggest an intrinsic theoretical error of
$\Delta L_{\rm th}\!=\!\val{6.8}{\MeV}$ [see Eq.~(\ref{R208vsL})], we expect a
cumulative error in the determination of $L$ of
\begin{equation}
 \Delta L = \sqrt{\Delta L_{\rm exp}^{2}+\Delta L_{\rm th}^{2}}
                         = \val{41.4}{\MeV}.
 \label{DeltaL}
\end{equation}
Can one improve on this limit? Clearly, if a follow-up to PREX-II is feasible
and the statistical accuracy of $\pm\val{0.02}{\fermi}$ is attained, then the error in $L$
could be significantly reduced to only $\Delta L=\val{15.2}{\MeV}$. 

Whereas PREX-II will place a stringent constraint on the density dependence of 
the symmetry energy (particularly on $L$), models predicting neutron radii of 
medium mass and light nuclei are affected by nuclear dynamics beyond $L$.
In particular, the Calcium Radius Experiment (CREX) will provide new and unique 
input into the isovector sector of nuclear theories and the high precision measurement 
of $R_n^{48}$ ($\pm\val{0.02}{\fermi}$) will help build a critical bridge between ab-initio 
approaches \cite{abinitio} and nuclear density functional theory \cite{dft_bender}. Note 
that CREX results can be directly compared to new coupled cluster calculations that 
are sensitive to three-neutron forces \cite{Hagen2012b,CREXworkshop}. Moreover, 
both \calcium[48] and \lead[208]---which are the basis of the approved CREX and 
PREX-II experiments\,\cite{ref:CREX}---are doubly-magic with a relatively simple 
nuclear structure, making them ideal candidates for accurate measurement of the
parity-violating asymmetry. In particular, the closed-shell nature of both nuclei 
results in a large energy gap (of a few MeV) to the first excited state. In combination 
with the high-resolution spectrometers at Jefferson Lab, this has the enormous 
advantage of ensuring that only the elastic electrons are captured in the detectors.

Finally, other nuclei such as \tin[120] and its heavier stable isotopes as well as 
measurements at other momentum transfers needed to map out the nuclear surface, should be 
considered. However, it appears that measurements on open-shell nuclei such
as \tin[] may not be as feasible or as cleanly interpretable as PREX and 
CREX\,\cite{Ban:2010wxNew}.   At present it is unclear whether any of
the stable \tin[]-isotopes may display a higher sensitivity to $L$ than \calcium[48] 
or \lead[208].  Thus, an important next step is to generate ``sensitivity plots'' akin to
Fig.~\ref{FigNSkins} for several semi-magic nuclei. Ultimately, the concordance
between CREX, PREX, and TREX (``Tin Radius EXperiment'') may provide the
best hope for a stringent constraint on the density dependence of the symmetry
energy.


\subsection{The way forward with giant resonances}
Besides the neutron-skin of heavy nuclei, the distribution of isoscalar monopole
strength has been found to be sensitive to the density dependence of the symmetry
energy. Unfortunately, this sensitivity is hindered by the relatively small neutron
excess of the stable nuclei explored to date. Thus, an important step forward is to
measure the distribution of monopole strength in exotic (very neutron-rich) nuclei
at next generation radioactive beam facilities. In particular, mapping GMR centroid
energies outside the stable \tin[112]--\tin[124] region is likely to provide
valuable insights. Moreover, such an experimental campaign may also help
elucidate the softness of Tin.

In the case of the isovector giant dipole resonance, the electric dipole polarizability
$\alphad$ was shown to be a strong isovector indicator. In particular, the combination
of  $\alphad^{208}$ times the symmetry energy at saturation density $J$ was seen to
be strongly correlated to the neutron-skin thickness $\RskinPb$.
 Although
photo-absorption experiments have been used for decades to probe the structure of
the GDR, it is critical to delve into the low-energy region for a proper account of
$\alphad$. Indeed, for the case of the exotic \nickel[68] nucleus, the PDR alone
accounts for about 25\% of $\alphad$. Here too the systematic exploration of the
isovector dipole strength along the chain of Sn-isotopes, for both stable and exotic
nuclei, may provide new insights into the problem.

\subsection{The way forward in theory}
An accurately calibrated  microscopic theory that both predicts and
provides well-quantified theoretical uncertainties is one of the goals of modern
nuclear energy density functionals~\cite{UNEDF,Kortelainen:2010hv}. The need to
quantify model uncertainties in an area such as theoretical nuclear physics is particularly
urgent as models that are fitted to experimental data are then used to extrapolate to the
extremes of density and isospin asymmetry. Ambitious theoretical programs aimed at
calibrating future EDFs using ground-state properties of finite nuclei, their collective
response, and neutron-star properties---supplemented by a proper covariance
analysis---are well on their way. A promising first step along these lines has already
been taken in Ref.~\citenum{Erler:2012qd}.

\subsection{The way forward in astrophysics}

There is now a small number of neutron stars with joint mass and radius constraints from spectral fitting.  
Constraining the EOS by a joint fit  is a powerful means of extracting information about dense matter.  To 
realize this potential will require the elimination of systematic errors in distance and spectral fitting.  With 
the anticipated parallax measurements from the \emph{Gaia} mission, the uncertainty in globular cluster 
distances will be drastically reduced.  Furthermore, the distances to nearby field neutron stars with identified 
companions, such as Cen X-4, will be known.  Nearby neutron stars, with a higher flux, would make a good 
target for spectral fitting.  Spectroscopy of the companion might also constrain the composition of the 
accreted material and therefore the emergent spectrum.
Improved distances would also help in the determination of mass and radius from X-ray bursts, if reliable 
models of the spectral evolution during the burst can be made.  Future timing missions such as 
\emph{LOFT} \citep{LOFT} and \emph{NICER} \citep{NICER} may provide further constraints on the 
mass-radius relation via rotation-resolved spectroscopy of pulsations.  It may also be possible to 
extract the compactness $M/R$ from observations of spectral edges observed in bursts with 
strong photospheric expansion \citep{Weinberg2005Exposing-the-Nu,t-Zand2010Evidence-of-hea}.

Further work should explore the implications of a small  (\val{9}{\kilo\meter}) neutron star radius.  
Such a small radius implies a softening of the symmetry energy for 
$\satdens \lesssim \dens \lesssim 2\satdens$, which can be probed by heavy ion collisions.  The 
softening needed for such small radii also suggests that the proton fraction remains low, 
and might preclude the onset of direct Urca neutrino cooling.  Further searches for cold, 
isolated neutron stars, as well as faint qLMXBs such as SAX~J1808.4$-$3658 and 1H~1905$+$00 
can inform us about the distribution of neutron stars with enhanced neutrino cooling.

Finally, the symmetry energy influences other quantities, such as the crust/core transition density 
and the fractional moment of inertia of the crust; it also affects the region of the crust with strongly 
deformed nuclei, the ``pasta'' phase.  These affect transport properties, such as the Ohmic resistivity 
of the neutron star crust, that might play a role in other observable phenomena 
\citep[see, e.g.,][]{arXiv:1304.6546}.

\subsection{The way forward for heavy ion collisions}

Heavy ion collisions provide the only means to study the density, isospin, and temperature dependence of the equation of state under controlled laboratory conditions.
The large dynamic range accessible with heavy ion collisions allows for the exploration of the
symmetry energy in very dilute matter
up to several times the saturation density. Specifically, the region at about twice saturation density is critical for the determination of neutron-star radii.
Placing strong constraints on the high-density component of the EOS of symmetric nuclear matter
represents an important achievement in studies of energetic heavy ion collision.
The community is now investigating the density dependence of the symmetry energy by
comparing neutron and proton observables, and in general isospin partners in light clusters, using
reactions with different neutron-proton asymmetries. 
Already a series of experiments of heavy ion collisions with energies in the Fermi energy range have constrained the symmetry energy from about 20\% above  $\satdens$ down to about $1/3\,\satdens$. These constraints agree well with, and in some cases even sharpen, the constraints from nuclear structure studies of exotic nuclei.  With the availability of facilities for the production of very asymmetric beams these possibilities are greatly enhanced.  The present focus is to improve experimental and theoretical uncertainties of the constraints at low density and on constraining the symmetry energy at supra-saturation densities, where it is rather poorly known from microscopic calculations, and where there are very few experimental data. The way forward will be to do new experiments that take advantage of studying reactions of wide asymmetries using  rare isotope beams to access a range of densities from below to well above saturation density with carefully chosen systems and observables which enhance the sensitivity to the symmetry energy.

Transport calculations are essential to extract the information about
the equation of state from the complex heavy ion collision process. By
considering appropriate observables, like ratios of isospin-sensitive
quantities, it has already been possible to obtain much information
about the symmetry energy. But it has also been learned that transport
calculations have to be refined to give a better description of
fluctuations, cluster and particle production, and generally have to
improve the consistency and reproducibility between different codes. A
planned series of workshops on code evaluation and comparison is an
important step in this direction.

\subsection{The way forward for dense QCD}
Quantum chromodynamics is presently  the ultimate theory of the strong interaction. As such, it determines the rich and complex structure of the phase diagram of strongly interacting matter. At low baryon density and high temperatures, collisions of heavy ions at facilities of increasing energy and sophistication are providing hints on the transition from hadronic matter to deconfined quark matter. At the opposite extreme, namely, high
densities and low temperatures, neutron stars offer the best (and perhaps unique) view of cold, dense matter.
The Facility for Antiproton and Ion Research (FAIR) in Europe provides the best alternative at bridging the
gap.  However, theoretical guidance into the detailed structure of the QCD phase diagram is hindered
by the intrinsically non-perturbative nature of QCD. Although much progress has been made, enormous
challenges remain. For example, QCD predicts that at ultra-high densities and low temperatures---where
the up, down, and strange quarks are effectively massless---the ground state is a superconductor with a
unique ``color-flavor locking" pairing scheme. Unfortunately, it appears that the enormous densities required
for this phase to develop can not be reached in the stellar cores. So the QCD ground state at the densities
of relevance to neutron stars remains an important challenge. On the other hand, lattice QCD provides
powerful insights into the nature of the phase  (or rather crossover) transition at finite temperature and zero
baryon density (or equivalently zero chemical potential $\mu$). While at present lattice simulations at
arbitrary values of the chemical potential are hindered by the vexing ``sign problem", renewed interest
in the QCD phase diagram has seen the emergence of various alternatives, including a Taylor series
(akin to the virial expansion) in the small parameter $\mu/T$. In particular, the main virtue of these exact
lattice results is that they provide critical benchmarks for assessing the reliability of other theoretical
approaches.

\begin{acknowledgments}
This work reports results from the first International Collaborations in Nuclear Theory (ICNT) program at NSCL/FRIB
during 2013.  We acknowledge the generous financial and logistic support from the director of NSCL/FRIB, which
made the ICNT program possible. The program was co-hosted by the Grants-in-Aid for Scientific
 Research on Innovative Areas (Nos.~24105001, 24105008, Area
 No.~2404) from JSPS.
 CJH acknowledges support from DOE grants DE-FG02-87ER40365 (Indiana
University) and DE-SC0008808 (NUCLEI SciDAC Collaboration).  EFB acknowledges support from NSF AST
grant 11-09176.  YK acknowledges support from the Rare Isotope Science Project funded by the
Ministry of Science, ICT and Future Planning (MSIP) and National Research Foundation
(NRF) of KOREA. MBT and WGL acknowledge the funding of NSF under Grant No. PHY-0606007 and DOE
Grant No.  DE-SC0004835. JP acknowledges support from DOE Grant No. DE-FD05-92ER40750.
HHW acknowledges support from the Excellence Cluster ``Origin and Structure of the Universe'' of the German Research Foundation (DFG).
\end{acknowledgments}

\bibliographystyle{apsrev}
\bibliography{bibs/Pasta5,bibs/cooling,bibs/ReferencesJP,bibs/HI,bibs/way_forward_YK}

\end{document}